\documentstyle[11pt]{scrartcl}
\input epsf
\def\paper#1#2#3#4#5{#1, {#3} {\bf #4}, \rm #5 (#2).}
\def\paperc#1#2#3#4#5#6{#1, {#3} {\bf #4}, \rm #5 (#2), {\tt #6}.}
\def\PRB{\it Phys. Rev. B}
\def\PRE{\it Phys. Rev. E}
\def\PRL{\it Phys. Rev. Lett}
\def\JPA{\it J. Phys. A}
\def\JPC{\it J. Phys. C}
\def\be{\begin{equation}}
\def\ee{\end{equation}}

\def\hlfplane{\frac 12\infty}
\def\opensquare{sq.}
\def\order{\mathop{O}\nolimits}
 
\def\simunder{\mathop{\sim }}

\begin{document}
\title{
  \begin{center}
    {\huge {\bf Phase transitions in two-dimensional random Potts models}}
  \end{center}
}

\author{Bertrand Berche and Christophe Chatelain\\[0.3cm]
       {\normalsize Laboratoire de Physique des Mat\'eriaux,}\\
       {\normalsize Universit\'e Henri Poincar\'e,  Nancy 1,}\\
       {\normalsize F-54506 Vand\oe uvre les Nancy Cedex, France}\\[0.5cm]}
\maketitle
\begin{abstract}
The influence of uncorrelated, quenched disorder on the phase transition of two 
dimensional Potts models will be reviewed.
After an introduction where the conditions of relevance of quenched randomness
on phase transitions are exemplified by  experimental measurements, the 
results of perturbative and numerical investigations in the case of the Potts 
model will be presented. The Potts model is of particular interest, since it 
can have in the pure case a second-order or a first-order transition, 
depending on the number of states per spin. In $2D$, transfer matrix 
calculations and Monte Carlo simulations were used in order to check the 
validity of conformal invariance methods in disordered systems. These 
techniques were then used to investigate the universality class of
the disordered Potts model, in both regimes of the pure model phase 
transitions. A test of replica symmetry became possible through a study of
multiscaling properties and a detailed analysis of the probability distribution
of the correlation functions was also made possible.
\end{abstract}
\newpage
\section{Introduction}\label{I}
Quenched disorder has
been the subject of an intensive activity in statistical physics 
during the past decades. 
The qualitative influence of 
disorder coupled to the energy-density at {\em second order} phase 
transitions is well understood since Harris proposed a celebrated
relevance criterion~\cite{Harris74}. 
At {\em first order} transitions, randomness obviously
softens the transitions, and, under some circumstances 
may even induce second 
order transitions according to a picture first proposed by Imry and 
Wortis~\cite{ImryWortis79} and then stated on more rigorous grounds 
by Aizenman 
and Wehr~\cite{AizenmanWehr89,HuiBerker89}, implying in particular that
an infinitesimal disorder induces  continuous transitions in $2D$.
Reviews can be found in the books of S.K. Ma, J.L. Cardy or Vik. 
Dotsenko~\cite{Ma76,Cardy96,Dotsenko01}.

\begin{figure}[ht]
  \vspace{8mm}
  \epsfysize=5.4cm
  \epsfbox{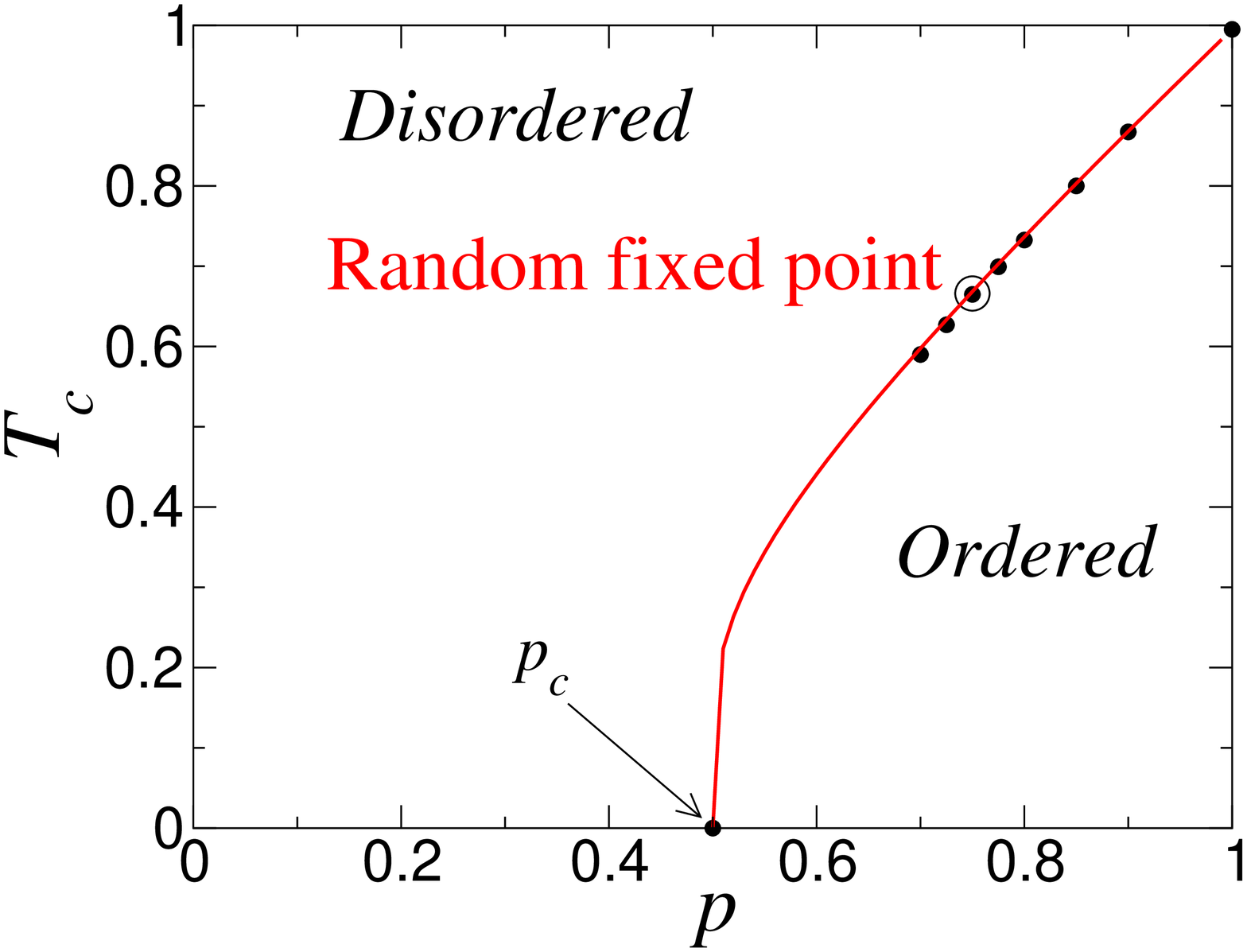}\hfill\ 
  \vskip -5.4cm
  \vskip.3mm\epsfysize=5.4cm
  \ \hfill\epsfbox{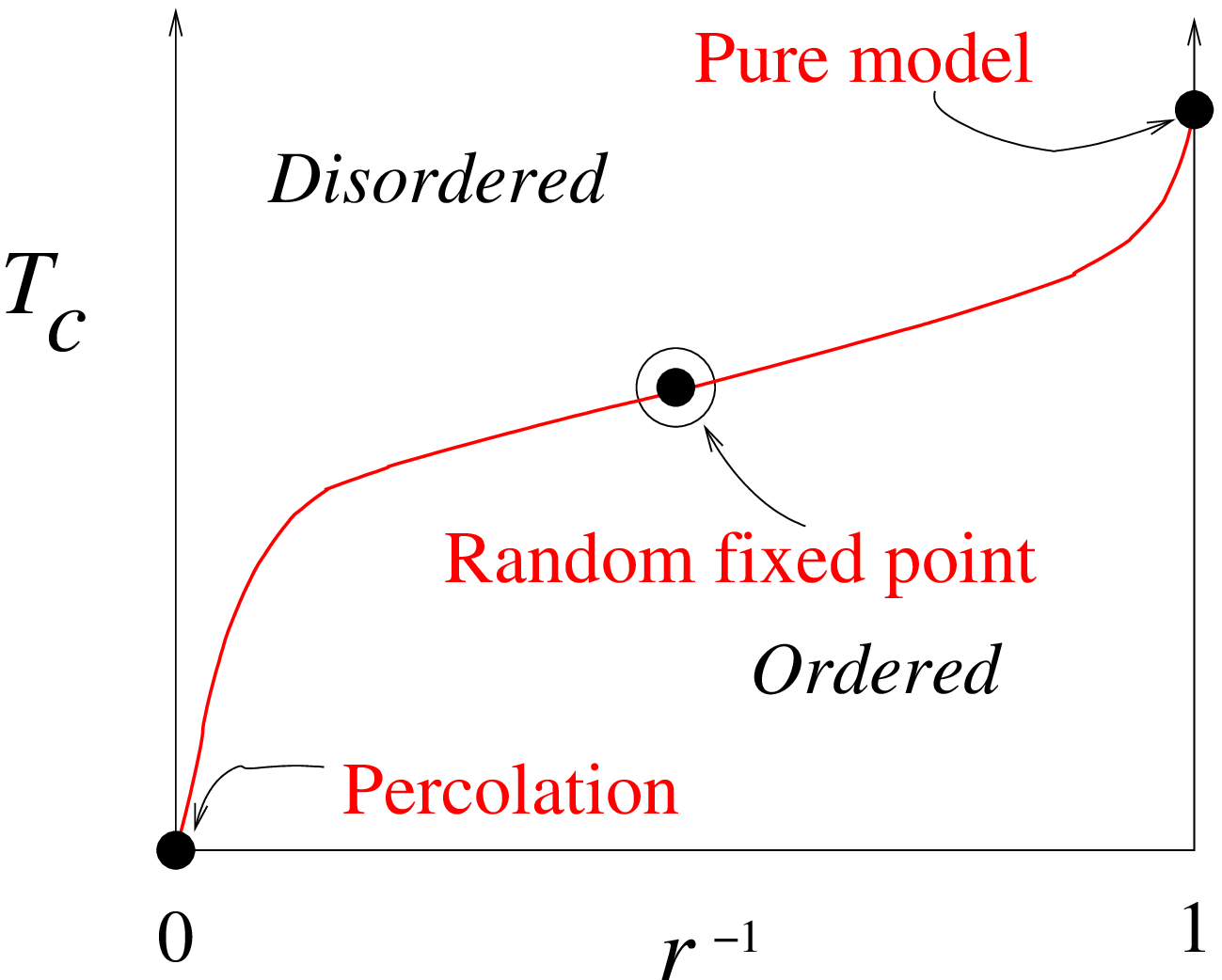} 
  \caption{Phase diagram of the dilution (left) and binary
    problems (right).} 
  \label{fig3}
\end{figure}

In spin models, the influence of quenched disorder strongly depends on
the nature of randomness, i.e. to which quantity the perturbation 
is coupled in the
Hamiltonian. Consider for example a very general model
with spins $\bf{s}_{\it i}$ on a lattice and
define the Hamiltonian
$$
-\beta{\cal H}=\sum_{(ij)}K_{ij}{\bf s}_i{\bf s}_j+\sum_i
{\bf H}_i{\bf s}_i+\sum_i D({\bf s}_i{\bf n}_i)^2+\dots
$$
where $K_{ij}$, ${\bf H}_i$, or ${\bf n}_i$ are independent random
quenched variables drawn from some probability distributions $P[K_{ij}]$,
$P[{\bf H}_i]$, or $P[{\bf n}_i]$, and which respectively describe
random-bond or dilute problems~\cite{Stinchcombe83}, 
random fields~\cite{Aharony78,ImryMa75,GrinsteinMa82,BricmontKupiainen87,ShapirAharony81}, 
and random anisotropy models~\cite{Aharony75,DudkaFolkHolovatch01}. 
As usually, the sum over $(ij)$ is supposed to be restricted 
to nearest neighbours.
Usually, we decide to work with uncorrelated quenched random variables,
for example $\overline{K_{ij}}\equiv\int K{\cal P}[K]{\rm d}K=K_0$, and $\overline{K_{ij}K_{kl}}=\Delta
\delta_{ik}\delta_{kl}$ 
and we will here only concentrate on the first
category, namely random-bond systems where disorder is coupled to
the energy density. Special cases of probability distributions of interest are
for example (all written here in the bond version of the problem)
\begin{description}
\item[i)] dilution problems, where non magnetic impurities are randomly
distributed  on the bonds or sites of the lattice, e.g. in the bond case
 \be
{\cal P}[K_{ij}]=\prod_{(ij)}
[p\delta(K_{ij}-K)+(1-p)\delta (K_{ij})],
\ee
\item[ii)] binary distributions, where we can imagine for example a
disordered alloy
of two magnetic species  
\be
{\cal P}[K_{ij}]=\prod_{(ij)}
[p\delta(K_{ij}-K)+(1-p)\delta (K_{ij}-Kr)],
\ee
\item[iii)] Gaussian distributions which are of particular interest
to perform Gaussian integration in analytic approaches, 
\be
{\cal P}[K_{ij}]=\prod_{(ij)}
[\frac{1}{\sqrt{2\pi\sigma^2}}\exp\left(-(K_{ij}-K)^2/2\sigma^2\right)]
\ee
\item[iv)] Continuous self-dual distributions (to be discussed later)
\be
{\cal P}[y_{ij}]=\prod_{(ij)}[(\cosh y_{ij}/\lambda)^{-1}],\ {\rm e}^{y_{ij}}=
\frac{1}{\sqrt q}({\rm e}^{K_{ij}}-1).
\label{eq5}\ee
\end{description}
In each case, there is a control parameter ($p$, $r$, $\sigma$ or $\lambda$)
which determines the strength of disorder.
The phase diagram is sketched in figure~\ref{fig3} for dilution and 
binary distribution. We expect a transition line between ordered and disordered
phases along which the transition is continuous in $2D$. For dilute problems,
there exists a percolation threshold where the transition temperature
vanishes and below which  long range order cannot
exist. In the bimodal case, the percolation fixed point at finite $q$ is reached in the limit
$r\to\infty$.  On the transition line, there should be some particular strength
of disorder corresponding to the location of the random fixed point,
where corrections to scaling should be small.

\begin{figure}[ht]
  \vskip.3mm\epsfxsize=11cm
  \centerline{\epsfbox{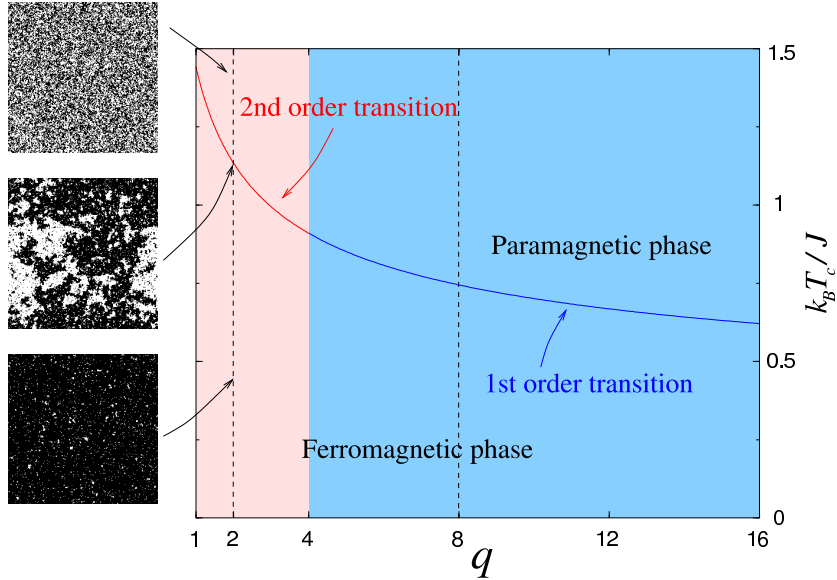}}
  \vskip 0truecm
  \caption{Phase diagram of the $2D$ pure Potts model as a function
          of the number of states per spin. The transition is of second-order
          when $q\le 4$ and becomes of first-order above. Inserts show Monte Carlo snapshots
        of typical spin configurations.} 
  \label{fig0} 
\end{figure}

The $q-$state Potts model~\cite{Wu82} is the 
natural candidate for the investigations of influence of disorder, 
since the pure model exhibits
two different regimes (see figure~\ref{fig0}): 
a second order phase transition when $q\leq 4$  and
a first order one for $q>4$ in two dimensions ($2D$). In $3D$, ordering
is easier and the transition becomes weakly first-order at $q=3$ already.

The 2-dimensional $q$-state Potts model is
defined by the 
following Hamiltonian :
        \begin{equation}
        -\beta{\cal H}=\sum_{(i,j)} K_{ij}\delta_{\sigma_i,\sigma_j}\label{eq-Ham0}
        \end{equation}
where the sum is restricted to nearest neighbours (here on a square lattice), the
degrees of freedom $\{\sigma_i\}$ can take $q$ values and the exchange
couplings $K_{ij}=J_{ij}/k_BT$ are quenched  independent random variables 
chosen according to some distribution $\{K_{ij}\}$  to be specified later.

Many results were obtained  for quenched randomness 
in this model in the last ten 
years, 
including 

\begin{description}
\item[i)] perturbative 
expansions for $2\le q \le 4$~\cite{Ludwig87,LudwigCardy87,Ludwig90,DotsenkoPiccoPujol95a,DotsenkoPiccoPujol95b,DotsenkoDotsenkoPiccoPujol95,DotsenkoPiccoPujol96,JugShalaev96,DotsenkoDotsenkoPicco97,Lewis98,JengLudwig01}. This work
was initiated in a series of papers by A.W.W. Ludwig and J.L. 
Cardy~\cite{Ludwig87,LudwigCardy87,Ludwig90}. 
A systematic study of energy-energy and spin-spin correlators in random bond 
Ising and Potts models, including tests of replica symmetry, was then performed
by a group around Vl.S. Dotsenko~\cite{DotsenkoPiccoPujol95a,DotsenkoPiccoPujol95b,DotsenkoDotsenkoPiccoPujol95,DotsenkoPiccoPujol96,JugShalaev96,DotsenkoDotsenkoPicco97,Lewis98}, 
with M. Picco, P. Pujol, Vik. Dotsenko,
J.L. Jacobsen, and M.-A. Lewis.
\item[ii)]
Monte Carlo 
simulations in both regimes $2\le q\le 4$ and 
$q>4$~\cite{Picco98,ChatelainBerche98a,ChatelainBerche98b,Yasaar98,ChatelainBerche99ccp,PalagyiChatelainBercheIgloi99,OlsonYoung99,ParedesValbuena99}.  
In the second order regime, simulations were first performed 
by S. Wiseman and E. Domany in the case
of the Ashkin-Teller model~\cite{WisemanDomany95} 
(which exhibits a critical line and for a given set of couplings 
belongs to the 
$4-$state Potts model universality class) and in the $3-$state Potts model by
J.K. Kim~\cite{Kim96}.
In the case $q=8$, the first 
Monte Carlo simulations were reported by S. Chen, A.M. Ferrenberg and 
D.P. Landau in 
Refs.~\cite{ChenFerrenbergLandau92,ChenFerrenbergLandau95}, but then in
both regimes, more accurate results were obtained by different groups,
like M. Picco~\cite{Picco98},  T. Olson and P. Young~\cite{OlsonYoung99}, 
or the present authors~\cite{ChatelainBerche98a,ChatelainBerche98b}. 
\item[iii)] Transfer 
matrices in both regimes~\cite{Glaus87,CardyJacobsen97,Picco97,JacobsenCardy98,ChatelainBerche99,JacobsenPicco00,ChatelainBerche00,Jacobsen00,ChatelainBerche01}. 
This technique was used to study random Potts models very early by
U. Glaus, then more refined computations were reported, for example
 by J.L. Cardy and J.L. Jacobsen~\cite{CardyJacobsen97,JacobsenCardy98} or in
Ref.~\cite{ChatelainBerche00}. 
\item[iv)] High-temperature series 
expansions, initially used in the random Ising 
model~\cite{RoderAdlerJanke98,RoderAdlerJanke99} to show the 
logarithmic corrections, then extended at $q=3$ in 
$3D$~\cite{HellmundJanke02a,HellmundJanke02b}, were also shortly applied to the
two-dimensional case~\cite{Janke02}.
\item[v)] Short-time dynamic 
scaling~\cite{YingHarada00,PanEtal01,YingEtal01,YingEtal02,YingEtal02bis}. 
\end{description}
We also notice
that dynamical properties (in order to discriminate between conventional
or activated dynamics) linked to non self-averaging have been recently
studied also~\cite{DeroulersYoung02}
and that the interesting limit $q\to\infty$ was
carefully investigated in Refs.~\cite{JacobsenPicco00,JuhaszeRiegerIgloi01}.
Reviews on selective parts of the subject were reported 
in Ph.D dissertations in
Refs.~\cite{Pujol96,Jacobsen98,Lewis99,Chatelain00,Davis01}.

Although closely related, the random-bond Ising model will not be discussed here, since
it was already the subject of many reviews, 
e.g.~\cite{DotsenkoDotsenko83,Shalaev94,SelkeShchurTalapov94}.
We will only remind here that in the pure $2D$ Ising model, the exponent $\alpha$
of the specific heat vanishes and
further investigation is needed, since the Harris criterion becomes inconclusive. It is now generally
believed that uncorrelated quenched randomness is a marginally irrelevant perturbation which 
does not modify the universal critical behaviour, but produces logarithmic corrections. For example the
singular part of the specific heat exhibits the following behaviour,
\begin{eqnarray}
&C_s(t)\sim\ln (1/{|t|})\qquad&\theta\ll{|t|}\ll 1,\nonumber\\
&C_s(t)\sim\ln\ln (1/{|t|})\quad&{|t|}\ll\theta ,\nonumber
\end{eqnarray}
$$\theta\sim{\rm e}^{-\pi /g^2},$$
where $g$ is the amplitude of disorder, linked to the impurity concentration. The behaviour of the
main physical quantities in the neighbourhood of the random fixed point is given in table~\ref{tab0}.
It was  confirmed unambiguously by Monte Carlo 
simulations~\cite{WangEtal90a,WangEtal90b,AndreichenkoEtal90c,TalapovShchur94} 
as shown in the same table.
We also note
that conformal mappings associated to Monte Carlo simulations
were initially used in the random-bond Ising problem by A. Talapov and
Vl.S. Dotsenko~\cite{TalapovDotsenko93}. 

{\small
\begin{table}[ht]
\begin{center}
\begin{tabular}{lll}
\hline
 \multicolumn{3}{c}{Random-bond Ising model} \\
 \cline{2-3}
 &\multicolumn{1}{c}{analytical results}& \multicolumn{1}{c}{numerical results}\\
\cline{1-1}\cline{2-2}\cline{3-3}
\hbox{\rm Correlation function}&$\langle\sigma_0\sigma_R\rangle
\sim R^{-1/4}(\ln R)^{1/8}$
&$\eta=0.2493\pm 0.0014$\\
\hbox{\rm Correlation length}&$\xi(t)
\sim {|t|}^{-1}[\ln (1/{|t|})]^{1/2}$
\\
\hbox{\rm Magnetisation}&$m(t)
\sim {|t|}^{1/8}[\ln (1/{|t|})]^{1/16}$
&${\beta}/{\nu}=0.1245\pm 0.0009$\\
\hbox{\rm Susceptibility}&$\chi(t)
\sim {|t|}^{-7/4}[\ln (1/{|t|})]^{7/8}$
&${\gamma}/{\nu}=1.7507\pm 0.0014$\\
\hbox{\rm Specific heat}&$C(t)
\sim \ln\ln (1/{|t|})$
\\
\hline
\end{tabular}
\end{center}
\caption{Critical behaviour of the random-bond Ising model.
\label{tab0}}
\end{table}
}

The numerical studies of disordered models showed that many difficulties, like
the lack of self 
averaging~\cite{Derrida84,AharonyHarris96,WisemanDomany98a,WisemanDomany98b} 
or varying effective exponents due to crossover phenomena. 
Averaging physical quantities over the samples 
with a poor statistics 
may thus lead to erroneous determinations of the critical exponents. 
Almost all the studies mentioned here were reported
in the case of the random bond system with self-dual probability distributions
of the coupling strengths in order to preserve the exact knowledge of the
transition line, which is an important simplification when one wants
to use finite-size-scaling techniques or conformal mappings which hold
at criticality only.

In real experiments on the other hand, disorder is inherent to the
working-out process and may result e.g. from the presence of impurities or
vacancies. For the description of such a disordered system, 
dilution is thus more realistic than for example a random distribution of
non-vanishing couplings (the so-called random-bond problem). 
Since universality is expected to hold, 
the detailed structure of the Hamiltonian
should not play any determining role in universal quantities like critical 
exponents,
but  crossover 
phenomena may alter the determination of the universality class.
Experimentally, the role of disorder in $2D$ systems has been investigated
in several systems.
Illustrating the influence of random defects in the case of the 
$2D$ Ising model universality class, samples made
of thin magnetic amorphous layers of 
(Tb$_{0.27}$Dy$_{0.73}$)$_{0.32}$Fe$_{0.68}$ of 10~\AA\  width, 
separated by non magnetic spacers of 100~\AA\  Nb in
order to decouple the magnetic layers were 
produced using sputtering techniques. A structural analysis (high resolution
transmission electron microscopy and $x$-ray analysis) was performed to
characterise the defects inherent to such amorphous structures, 
and in spite of these random defects 
separated on average by a distance of a few nm, the samples were 
shown to exhibit Ising-like 
singularities with critical exponents~\cite{MohanEtal98} 
$$ \beta=0.126(20),\ \gamma=1.75(3),\ \delta=15.1(10).$$
This is coherent with the fact that disorder does not change the universality
class of the $2D$ Ising model, apart from logarithmic corrections
which are probably impossible to observe experimentally, since their role 
becomes prominent only in the very neighbourhood of the critical point.
A  beautiful experimental confirmation 
of the Harris criterion -- which predicts a modification of the critical
behaviour in random systems when 
the exponent $\alpha$ of the 
specific heat is positive for the pure system --
was reported in a Low Energy Electron Diffraction  
investigation of a 
$2D$ order disorder 
transition~\cite{SchwengerEtAl94} belonging to the
4-state Potts model universality class. Order-disorder transitions of 
adsorbed atomic layers are known to belong to different two-dimensional 
universality classes depending on the type of superstructures in the ordered 
phase of the ad-layer~\cite{DomanyEtAl78,Rottman81}. 
The substrate plays a major role in ad-atom ordering,
as well as the coverage (defined as the number of ad-atoms per surface atom)
which determines the possible superstructures of the over-layer.
For example, sulfur chemisorbed on 
Ru$(001)$ exhibits four-state or three-state Potts critical singularities
for the $p(2\times 2)$ and the $(\sqrt 3\times\sqrt 3)R30^\circ$
respectively~\cite{SokolowskiPfnur94} (at coverages $1/4$ and $1/2$). 
The case of the $(2\times 2)-$2H/Ni$(111)$ order-disorder transition of
hydrogen adsorbed on the $(111)$ surface of Ni thus belongs to the
$2D$ four-state Potts model universality class, since the ground state,
stable at low temperatures, has a four-fold degenaracy due to the 
four possible coverings of the ad-atoms at the $(111)$ surface
(see figure~\ref{fig1}).

\begingroup
\begin{figure}[ht]
  \vskip.3mm\epsfysize=5cm
  \centerline{\epsfbox{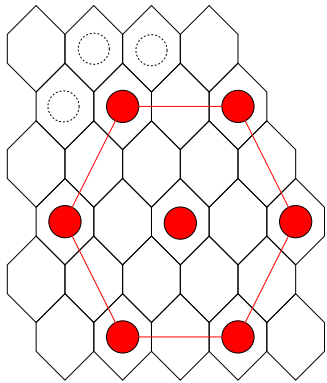}}
  \vskip 0truecm
  \caption{$(2\times 2)-$2H/Ni$(111)$ order-disorder transition of
          hydrogen. The ground-state has a four-fold degenaracy due to the 
          four possible covering of the $(111)$ surface of Ni by H ad-atoms.} 
  \label{fig1}
\end{figure}
\begin{figure}[ht]
  \vskip.3mm\epsfysize=5cm
  \centerline{\epsfbox{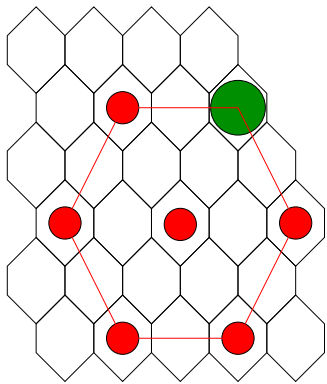}}
  \vskip 0truecm
  \caption{$(2\times 2)-$2H/Ni$(111)$ order-disorder transition of
          hydrogen with oxygen impurities randomly 
          chemisorbed on the surface, and occupying some of the lattice
          sites.} 
  \label{fig2}
\end{figure}
\endgroup

The expected exponents are thus close to theoretical values of
$\beta=1/12\simeq 0.083$, $\gamma=7/6\simeq 1.167$, and $\nu=2/3\simeq 0.667$
for example. 

Low energy electron diffraction (LEED) makes possible 
to measure these exponents through the diffracted intensity
$I({\bf q})$ or structure factor. This is the two-dimensional Fourier transform
of the pair correlation function of ad-atom density. Long range 
fluctuations produce an isotropic Lorentzian 
centered at the superstructure 
spot position  ${\bf q}_0$ with a peak intensity given by the
susceptibility and a width determined by the inverse correlation length, while
long range order gives a background signal  proportional to the order 
parameter squared:
$$        I({\bf q})=\langle m^2\rangle\delta({\bf q}-{\bf q}_0)
        +\frac{\chi}{1+\xi^2({\bf q}-{\bf q}_0)^2}.
$$
The following exponents were thus 
measured~\cite{SchwengerEtAl94,BuddeEtAl95,VogesPfnur98}
$$\beta=0.11\pm 0.01,\ \gamma=1.2\pm 0.1,\ \nu=0.68\pm 0.05$$ 
in correct 
agreement with 4-state Potts values (the small deviation, especially for the
exponent $\beta$, is attributed to the logarithmic corrections
to scaling of the pure 4-state Potts model~\cite{CardyNauenbergScalapino80}).
The same experiments were then reproduced in the presence
of  intentionally added oxygen
impurities, at 
a temperature which is above the ordering temperature of pure oxygen
adsorbed on the same substrate. The mobility of these oxygen atoms is
furthermore considered to be low enough at the hydrogen order-disorder
transition critical temperature that they essentially represent quenched
impurities randomly distributed in the hydrogen layer 
(see figure~\ref{fig2}). The 
exponents become 
$$
\beta=0.135\pm 0.010,\  \gamma=1.68\pm 0.15,\  
\nu=1.03\pm 0.08,
$$ 
which definitely supports the modification of the universality class
in the presence of quenched disorder, in agreement with Harris
criterion ($\alpha=2/3$ for the 4-state Potts model).

The aim of this lecture is to provide a review of both perturbative and 
numerical studies of the bond diluted Potts model for several values 
of the number of states per spin (in order to cover the two
different regimes of the pure system's phase transitions).
In section~\ref{sec:two}, the perturbative approach is discussed and
the essential results are summarised in the $2D$ case with $2\le q\le 4$.
The details of the numerical techniques used in two dimensions 
are presented in section~\ref{sec:three} and ~\ref{sec:four}, 
while the comparison between numerical and analytical results 
is made in ~\ref{sec:five}.

\section{Perturbative approach in $2D$}\label{sec:two}
\subsection{Replicas and relevance criterion}
For a specific disorder realization $[K_{ij}]$, the Hamiltonian
${\cal H}[K_{ij},\sigma_i]$ is written
\begin{equation}
  -\beta{\cal H}[K_{ij},\sigma_i]=\sum_{(ij)}(K_0+\delta K_{ij})
  \delta_{\sigma_i,\sigma_j}.
\end{equation}
The corresponding partition function and free energy are given as follows
$$
Z[K_{ij}]=\int{\cal D}[\sigma_i]{\rm e}^{-\beta{\cal H}[K_{ij},
\sigma_i]},$$
$$
F[K_{ij}]=-k_BT\ln Z[K_{ij}].$$
To get the quantities of interest, for example the average free energy, 
we have to perform an average over the distribution 
${\cal P}[K_{ij}]$\footnote{In
the case of an annealed disorder, the impurities are thermalized (this
would only be possible if the relaxation time of randomness is small
compared to the time scale of the experiment) and their
probability distribution ${\cal P}[K_{ij}]$ depends strongly on the spins
(and vice-versa); since it is the equilibrium distribution
${\cal P}[K_{ij}]=\int{\cal D}[\sigma_i] Z^{-1}{\rm e}^{-\beta{\cal H}[K_{ij},
\sigma_i]}$, where $Z$ does no longer depend on the disorder
realization, but is obtained through
$Z=\int{\cal D}[K_{ij}]{\cal D}[\sigma_i] {\rm e}^{-\beta{\cal H}[K_{ij},
\sigma_i]}$. In the annealed case, if the impurity concentration is
kept constant, there is a `Fisher's renormalization' of the exponents if
the specific heat of the pure system is diverging~\cite{Fisher68}.
},
 \begin{equation}
   F=\overline{F[K_{ij}]}
   =-k_BT\int {\cal D}[K_{ij}] {\cal P}[K_{ij}]\ln Z[K_{ij}].
\end{equation}
Averaging the logarithm of the partition function is possible through the 
identity
$$\ln Z=\lim_{n\to\infty}\frac{1}{n}(Z^n-1),$$
which requires, before averaging, to produce $n$ copies (with labels
$\alpha$) of the system with the same disorder configuration,
$$
(Z[K_{ij}])^n=\int\left(\prod_{\alpha=1}^n{\cal D}[{\sigma}^{(\alpha)}_i]\right)
{\rm e}^{-\beta\sum_\alpha{\cal H}[K_{ij},
{\sigma}^{(\alpha)}_i]},$$
and then to perform integrations over  ${\cal P}[K_{ij}]$,
$$
\overline{{\rm e}^{-X}}={\rm e}^{-\bar X+\frac{1}{2}(\overline{X^2}-
\bar X^2)+\dots},$$
leading to
\begin{eqnarray}
\overline{(Z[K_{ij}])^n}&=&
\int\left(\prod_{\alpha=1}^n{\cal D}[{\sigma}^{(\alpha)}_i]\right)
{\rm e}^{-\sum_\alpha
(K_0+\overline{\delta K})\sum_{(ij)}
\delta_{\sigma_i^{(\alpha)},\sigma_j^{(\alpha)}}}\nonumber\\
&&\times{\rm e}^{\sum_{\alpha\not=\beta}
(\overline{\delta K^2}-\overline{\delta K}^2)\sum_{(ij)}  
\delta_{\sigma_i^{(\alpha)},\sigma_j^{(\alpha)}}
\delta_{\sigma_i^{(\beta)},\sigma_j^{(\beta)}}
+\dots}\label{Zrepliques}
\end{eqnarray}
In the leading term, $\overline{\delta K_{ij}}$ has RG eigenvalue
$y_t=d-x_\varepsilon$ and corresponds to a simple shift of the transition 4
temperature (which is obviously a relevant effect). In the next term,
the second moment of the distribution,
$\overline{\delta K^2}-\overline{\delta K}^2$, has RG eigenvalue
$y_H=d-2x_\varepsilon$, and all the following terms are 
irrelevant\footnote{The leading (unperturbed) term is written in the continuum
limit as 
$-\beta{\cal H}_c=m_0\int\sum_\alpha
\varepsilon_\alpha({\bf r}){\rm d}^2{\bf r}$ where $m_0$ stands for
$K_0+\overline{\delta K}$
while the perturbation is written
$g_0\int\sum_{\alpha\not=\beta}
\varepsilon_\alpha({\bf r})\varepsilon_\beta({\bf r}){\rm d}^2{\bf r}$
with $g_0$ corresponding to $\overline{\delta K^2}-
\overline{\delta K}^2$.}.
Using hyperscaling relation, the Harris scaling dimension of disorder
is rewritten
\be
y_H=\alpha/\nu.\ee
It implies that at second-order transitions, disorder is a relevant
perturbation which modifies the critical behaviour when the
specific heat  exponent $\alpha$ of the pure system is positive, while
it is irrelevant (and universal properties are thus unaffected by randomness)
when $\alpha$ is negative. In the borderline case $\alpha=0$, randomness
is marginal to leading order. This is the case for example of the $2D$
Ising model discussed in the introduction, 
where quenched disorder is eventually marginally irrelevant
and produces only logarithmic corrections to the unchanged leading
critical behaviour.

The case of first-order transitions was considered later, by Imry and
Wortis, Aizenman and Wehr, then Hui and 
Berker~\cite{ImryWortis79,AizenmanWehr89,HuiBerker89}. It can be 
intuitively understood from the above results simply by
noticing that the existence of a latent heat at first-order transition
corresponds to a discontinuity of the energy density and
can be described by a vanishing energy density scaling dimension, so that
disorder is always relevant in this sense.

\subsection{Perturbation techniques}
\subsubsection{Average correlation functions}
Many results were obtained in the $2D$ random Potts model using RG 
perturbations, mainly around Ludwig and Vl. Dotsenko.  
In equation~(\ref{Zrepliques}), it appears that $\overline{Z^n}$ couples
the replicas via energy-energy interactions
$\sum_{\alpha\not=\beta}
(\overline{\delta K^2}-\overline{\delta K}^2)\sum_{\bf r}
\varepsilon_\alpha({\bf r})\varepsilon_\beta({\bf r})$
 which have to be treated
as a perturbation around the pure fixed point. 
Here, $\varepsilon_\alpha({\bf r})$ is a short notation for 
$\delta_{\sigma_i^{(\alpha)}\sigma_j^{(\alpha)}}$, and ${\bf r}$ stands for
the lattice  unit vectors. The second  cumulant of the coupling
 distribution
 will be denoted  $g_0$ in the following.
 Two different  schemes
 were considered  in the literature~\cite{DotsenkoDotsenkoPicco97,Pujol96}, 
\begin{description}
\item[i)] replica symmetric scenario, where all the replicas are
coupled through the same interaction strength,
\be\sum_{\alpha\not=\beta}
g_0\sum_{\bf r}
\varepsilon_\alpha({\bf r})\varepsilon_\beta({\bf r}),
\ee 
\item[ii)] replica symmetry breaking scenario, where the coupling
between replicas are
replica-dependent,
\be\sum_{\alpha\not=\beta}
g_{\alpha\beta}\sum_{\bf r}
\varepsilon_\alpha({\bf r})\varepsilon_\beta({\bf r}).
\ee 
\end{description}
The program is thus to consider $2D$ Potts model with weak bond
randomness, compute the scaling dimensions $x'_\sigma(n)$ of the order 
parameter and $x'_\varepsilon(n)$ of the energy-density
perturbatively\footnote{The primes denote the scaling dimensions at the random fixed point.} 
around the Ising 
model conformal field theory, and then take the replica limit $n\to 0$.
Expansions are performed around the pure model fixed point (weak disorder)
in terms of the disorder strength 
$\overline{\delta K_{ij}^2}-\overline{\delta K_{ij}}^2$, and the exponents
are given in powers of $y_H=\alpha/\nu$.

Different expansion parameters can be found in the literature, and it
is worth collecting the main notations. 
The Potts models can be identified to minimal conformal 
models~\cite{FriedanQiuShenker84}
which are
parametrised by an integer $m$ which determines the central charge and
critical behaviour of the model. The correspondence is given by
$$m=\frac{\pi}{\cos^{-1}(\sqrt q/2)}-1$$
and the central charge and exponents follow from
\begin{eqnarray} 
c&=&1-\frac{6}{m(m+1)},\nonumber\\
x_\varepsilon&=&\frac{m+3}{2m},\nonumber\\
x_\sigma&=&\frac{(m+3)(m-1)}{8m(m+1)}.\nonumber
\end{eqnarray}
We note that $m=3$ for the Ising model, $m=5$ for the $3-$state
Potts model and $m\to\infty$ for the $4-$state Potts model.
 The Harris RG eigenvalue becomes
$$ y_H=\frac{m-3}{m},$$ 
which is
proportional to $q-2$ to linear order
($q$ being the number of states per spin of the
Potts model):
\be
y_H=\frac{4}{3\pi}(q-2)-\frac{4}{9\pi^2}(q-2)^2+O[(q-2)^3].
\ee
Using a Coulomb gas representation, a natural expansion parameter 
$\epsilon$ is defined through
$$ \alpha_+^2=\frac{m+1}{m}=\frac 43+\epsilon,$$
and it is linked to $y_H$ by
$\epsilon=-\frac 13y_H$.

\begin{description}
\item[i)] The replica symmetric case is based on the assumption that replica
symmetry is not broken initially, and is then preserved by the 
renormalization group.
The renormalization of the coupling constant $g_0$ 
is determined by perturbative
calculation using the operator algebra. 
For any scaling operator $\phi$, the perturbed two-point correlation function
$\langle \phi(0)\phi({\bf R})\rangle_g$ corresponds, in the limit
$n\to 0$,
to the average correlator $\overline{\langle \phi(0)\phi({\bf R})\rangle}$.
We can write
$$
\langle \phi(0)\phi({\bf R})\rangle_g=
\frac{{\rm Tr}{\ \!\phi(0)\phi({\bf R})\rm e}^{-\beta({\cal H}_c+{\cal H}_g)}
}
{{\rm Tr}{\ \!\rm e}^{-\beta({\cal H}_c+{\cal H}_g)}}
$$
where the perturbation term $-\beta{\cal H}_g=g_0
\int\sum_{\alpha\not=\beta}
\varepsilon_\alpha({\bf r})\varepsilon_\beta({\bf r}){\rm d}^2r$
acts on the `critical' Hamiltonian
$-\beta{\cal H}_c=m_0
\int\sum_{\alpha}
\varepsilon_\alpha({\bf r}){\rm d}^2r+
h_0
\int\sum_{\alpha}
\sigma_\alpha({\bf r}){\rm d}^2r$, the last term being included in order to
compute the renormalization of the spin operator.

When expanded in terms of unperturbed correlators, it yields the following
expansion~\cite{PatashinskiiProkowskii79,Polyakov73}
\begin{eqnarray}
\langle \phi(0)\phi({\bf R})\rangle_g&=&
\langle \phi(0)\phi({\bf R})\rangle_0
-\beta\langle{\cal H}_g \phi(0)\phi({\bf R})\rangle_0\nonumber\\
&&+\frac 12 \beta^2\langle{\cal H}_g^2 \phi(0)\phi({\bf R})\rangle_0
+\dots\nonumber
\end{eqnarray}
The renormalization of the coupling constant follows from the 
expansion~\cite{DotsenkoPiccoPujol95a,DotsenkoPiccoPujol95b,DotsenkoDotsenkoPiccoPujol95,DotsenkoPiccoPujol96}
\begin{eqnarray}
&&g_0
\int\sum_{\alpha\not=\beta}
\varepsilon_\alpha({\bf r})\varepsilon_\beta({\bf r}){\rm d}^2r
\nonumber\\
&&+\frac 12 g_0^2
\int\sum_{\alpha\not=\beta}
\varepsilon_\alpha({\bf r})\varepsilon_\beta({\bf r}){\rm d}^2r
\int\sum_{\gamma\not=\delta}
\varepsilon_\gamma({\bf r}')\varepsilon_\delta({\bf r}'){\rm d}^2r'+\dots
\nonumber\\
&&=g
\int\sum_{\alpha\not=\beta}
\varepsilon_\alpha({\bf r})\varepsilon_\beta({\bf r}){\rm d}^2r,
\end{eqnarray}
leading to $g=g_0(1+A_1g_0+A_2g_0^2+\dots)$. The successive terms are
obtained from operator product expansions. In the first order term, the
dominant contribution to $A_1$ 
follows from contraction of neighbouring pairs, $\varepsilon({\bf r})
\varepsilon({\bf r}')\sim |{\bf r}-{\bf r}'|^{-2x_\varepsilon}$, 
in the same replica ($\beta=\gamma\not =\alpha,\delta$ and 
$\alpha\not=\delta$). Such an expression has to be understood inside 
unperturbed correlators.
Including combinatorial factors (they are $2(n-2)$ such
factors), integration over space up to an infrared cutoff, 
$b> |{\bf r}-{\bf r}'|$, leads to
$\frac 12 g_0^2A_1
\int\sum_{\alpha\not=\delta}
\varepsilon_\alpha({\bf r})\varepsilon_\delta({\bf r}){\rm d}^2r$,
where $A_1$ is dominated by
$$A_1=2(n-2)\int_{|{\bf r}-{\bf r}'|<b}|{\bf r}-{\bf r}'|^{-2x_\varepsilon}
{\rm d}^2r'=4\pi(n-2)\frac{b^{2-2x_\varepsilon}}{2-2x_\varepsilon}.$$
Since $y_H=2-2x_\varepsilon$, one recovers the Harris criterion.
Up to the first order, we get the following expression for $g$ in terms of the
bare coupling constant $g_0$, $g=g_0(1+2\pi (n-2)\frac{1}{y_H}b^{y_H}g_0+
\order(g_0^2))$. Following Dotsenko and co-workers,
the coupling constants are multiplied by a factor $b^{y_H}$ in order to get
dimensionless coupling constants $g(b)$.

The $\beta-$function 
up to second order in the
$n\to 0$ limit is finally given by
\be \beta(g)=\frac{{\rm d}g(b)}
{{\rm d}\ln b}=y_Hg(b)-8\pi g^2(b)+32\pi^2 g^3(b)
+\order (g^4(b)).\ee
It leads to a non trivial IR fixed point (which determines the long distance
physics) $g_c=\frac{1}{8\pi}y_H+\frac{1}{16\pi}y_H^2+\order (y_H^3)$.

Renormalization of the energy  and the order parameter density operators
follow from the same analysis, e.g.:
\begin{eqnarray}
&&m_0
\int\sum_{\alpha}
\varepsilon_\alpha({\bf r}){\rm d}^2r\left(1
+g_0\int\sum_{\beta\not=\gamma}
\varepsilon_\beta({\bf r}')\varepsilon_\gamma({\bf r}'){\rm d}^2r'\right.
\nonumber\\
&&\left.+\frac 12g_0^2
\int\sum_{\beta\not=\gamma}
\varepsilon_\beta({\bf r}')\varepsilon_\gamma({\bf r}'){\rm d}^2r'
\int\sum_{\delta\not=\eta}
\varepsilon_\delta({\bf r}'')\varepsilon_\eta({\bf r}''){\rm d}^2r''
+\dots\right)
\nonumber\\
&&=m
\int\sum_{\alpha}
\varepsilon_\alpha({\bf r}){\rm d}^2r,
\end{eqnarray}
and provide the expansions (details of the calculation of the integrals,
using a Coulomb gas representation, can be found e.g. in 
Ref.~\cite{DotsenkoPiccoPujol95a})
$m=m_0(1+B_1g_0+B_2g_0^2+\dots)=Z_\varepsilon m_0$ and
$h=h_0(1+C_1g_0+C_2g_0^2+\dots)=Z_\sigma h_0$, leading when $n\to 0$ to
\begin{eqnarray} 
\gamma_\varepsilon(g)&=&\frac{{\rm d}\ln Z_\varepsilon}{{\rm d}\ln b}
=-4\pi g(b)+8\pi^2g^2(b),\nonumber\\
\gamma_\sigma(g)&=&\frac{{\rm d}\ln Z_\sigma}{{\rm d}\ln b}
=-\pi^2y_Hg^2(b)\left(
1+2
\frac{\Gamma^2(-{\scriptstyle\frac 23})
\Gamma^2({\scriptstyle\frac 16})}{\Gamma^2(-{\scriptstyle\frac 13})
\Gamma^2(-{\scriptstyle\frac 16})}
\right)+8\pi^2g^3(b).\nonumber\\
\end{eqnarray}

For the correlators themselves, the renormalization  equations can be written
$$
\langle \phi(0)\phi(s{\bf R})\rangle
=\frac{Z^2_\phi(g(bs))}{Z^2_\phi(g(b))}s^{-2x_\phi}
\langle \phi(0)\phi({\bf R})\rangle,
$$
where $s$ is the scaling factor.
Using now
$$
\gamma_\phi(g)=\frac{{\rm d}\ln Z_\phi}{{\rm d}\ln b},
$$
or $\ln Z_\phi=\int\gamma_\phi(g){\rm d}\ln b$,
the ratio $\frac{Z^2_\phi(g(bs))}{Z^2_\phi(g(b))}$ can be rewritten
$$
\frac{Z_\phi^2(bs)}{Z_\phi^2(b)}={\rm e}^
{2\int_b^{bs}\gamma_\phi(g){\rm d}\ln b}
$$
which is dominated at long distances by $g\simeq g_c$, such that 
$\int_b^{bs}\gamma_\phi(g){\rm d}\ln b\simeq \gamma_\phi(g_c)\ln s$.
The homogeneity equation  thus becomes
$$
\langle \phi(0)\phi(s{\bf R})\rangle
= s^{-2(x_\phi-\gamma_\phi(g_c))}\langle \phi(0)\phi({\bf R})\rangle,
$$
and choosing a rescaling factor $s=R^{-1}$, the  two point correlator
decays as
\be
\langle \phi(0)\phi({\bf R})\rangle
\simeq R^{-2(x_\phi-\gamma_\phi(g_c))}.
\ee
The corresponding scaling dimension is modified according to
\be
x'_\phi =x_\phi-\gamma_\phi(g_c).\ee

Collecting the results of Dotsenko and co-workers, we give the
new thermal and magnetic scaling dimensions (with primes) in terms of
the original ones 
(unprimed)~\cite{DotsenkoPiccoPujol95a,DotsenkoPiccoPujol95b,DotsenkoDotsenkoPiccoPujol95,DotsenkoPiccoPujol96}:
\begin{eqnarray}
x'_\varepsilon&=&x_\varepsilon-\gamma_\varepsilon(g_c)\nonumber\\
&=&x_\varepsilon+\frac 12 y_H+\frac 18 y_H^2+\order (y_H^3)\\
x'_\sigma&=&x_\sigma-\gamma_\sigma(g_c)\nonumber\\
&=&x_\sigma+\frac{1}{32}\frac{\Gamma^2(-{\scriptstyle\frac 23})
\Gamma^2({\scriptstyle\frac 16})}{\Gamma^2(-{\scriptstyle\frac 13})
\Gamma^2(-{\scriptstyle\frac 16})} y_H^3+\order (y_H^4)
\label{eq:x_s-dotsenko}
\end{eqnarray}

\item[ii)] The other assumption of a broken replica symmetry leads to a
different fixed point structure.
The coupling between replicas, $g_{\alpha\beta}$, is now dependent on
the pair indexes, and it is generalised to a continuous variable $x$
instead of pair indices, $g_{\alpha\beta}\to g(\alpha-\beta)=g(x)$.
It is found that there is only one marginally attractive solution for the 
coupling $g(x)$ which then enable to compute $\gamma_\varepsilon (g)$
and $\gamma_\sigma (g)$, leading to a modified thermal exponent
\be
x''_\varepsilon=x_\varepsilon+\frac 12y_H+\order (y_H^3),
\ee
while to $y_H^3$ order, the magnetic scaling index remains
the same as in the replica symmetric 
scenario~\cite{DotsenkoDotsenkoPiccoPujol95}.
\end{description}

\subsubsection{Multiscaling and higher order moments of the correlators}
In order to measure some other differences between the replica symmetric and 
the replica symmetry breaking cases, the moments of the correlators can
also be helpful. For any scaling field $\phi({\bf r})$, multiscaling arises
when the scaling dimensions associated to the moments of the correlators
do not follow a simple linear law:
\be
\overline{
\langle\phi(0)\phi({\bf R})\rangle^p}\sim R^{-2px_{\phi^p}},\ 
x_{\phi^p}\not = x_{\phi}.\ee
In the magnetic sector, a difference between the two cases 
occurs to $y_H^2$ order for the
second moment~\cite{DotsenkoDotsenkoPicco97}:
\begin{eqnarray}
&RS&x'_{\sigma^2}=x_\sigma-\frac{1}{16}y_H+\frac{1}{32}\left( 4\ln 2-
{\scriptstyle\frac{11}{12}}\right)y_H^2+\order (y_H^3),\\
&RSB&x''_{\sigma^2}=x_\sigma-\frac{1}{16}y_H+\frac{1}{32}\left( 4\ln 2-
{\scriptstyle\frac{5}{12}}\right)y_H^2+\order (y_H^3).
\label{eq:RS}\end{eqnarray}
For higher order moments, the computation were only performed in the replica
symmetric case~\cite{Ludwig90,Lewis98,JengLudwig01}, leading to
\begin{eqnarray}
x'_{\varepsilon^p}&=&1-\frac{2}{9\pi^2}(3p-4)(q-2)^2+O[(q-2)^3],\\
x'_{\sigma^p}&=&1-\frac{1}{16}(p-1)y_H^2\nonumber\\
&&-\frac {1}{32}(p-1)
[{\scriptstyle\frac{11}{12}-4\ln 2+\frac{1}{24}(33-29\sqrt 3\pi/3)(p-2)}]
y_H^3+\order (y_H^4).
\label{eq:x_lewis}\end{eqnarray}
\subsubsection{Are these effects measurable?}
The question is now to try to detect numerically the effects discussed above.
These are perturbative expansions around $q=2$, so that a natural choice of
system to measure the scaling dimensions in the presence of quenched disorder
is the $3-$state Potts model. At $q=3$ we have $x_\varepsilon=4/5$ and
$y_H=2/5$ from which the perturbed  scaling dimensions in the energy and
magnetic sectors can be obtained. The values are given in table~\ref{tab1}.
The numerical data clearly show that the expected variations are quite
small and need accurate numerical techniques to discriminate between
RS and RSB schemes.

\begin{table}[ht]
\begin{center}
\begin{tabular}{llllll}
\hline
Scheme  & \multicolumn{5}{c}{Scaling dimensions} \\
  \cline{2-6} 
 & $x_\sigma$ & $x_\varepsilon$ & $x_{\sigma^2}$ & $x_{\sigma^0}$ & 
$x_{\varepsilon^0}$ \\ 
\hline
Pure system & 0.13333 & 0.800 & 0.13333 & 0.13333 & 0.800\\
RS & 0.13465 & 1.000 & 0.11761 & 0.18303 & 1.090 \\
RSB & 0.13465 & 1.020 & 0.12011 & -- & -- \\
\hline
\end{tabular}
\end{center}
\caption{Comparison between pure $3-$state Potts model critical exponents
  and the expected values obtained from perturbation expansions. The
  notation $x_{\sigma^2}$
  corresponds to the second moment of the spin-spin correlation function, while
  $x_{\sigma^0}$ and $x_{\varepsilon^0}$ are associated to the typical 
  correlations.
\label{tab1}}
\end{table}

\newpage

\section{Numerical techniques in $2D$\label{sec:three}}
\subsection{Monte Carlo simulations}
\subsubsection{Cluster algorithms}
For the simulation of spin systems, standard Metropolis algorithms based
on local updates of single spins suffer from the well known 
critical slowing down. As the second-order phase transition is approached, the
correlation length becomes 
larger and the system contains larger and larger clusters
in which all the spins are in the same state. Statistically independent
configurations can be obtained by local iteration rules only after a long
dynamical evolution which needs a huge number of MC steps. This makes this
type of algorithm inefficient close to a critical point. 

Since the transition of the disordered Potts model is always
expected to be on average a second-order one, the resort
to cluster update 
algorithms is more convenient~\cite{Janke96,BarkemaNewmann97}.
The main recipe of cluster algorithms is the identification of 
clusters of sites using a bond percolation process connected to the spin
configuration. The spins of the clusters are then independently flipped.
A cluster algorithm is particularly efficient if the percolation 
threshold coincides with the transition point of the spin model, 
which guarantees that clusters of all sizes will be updated in a single
MC sweep.

In the case of the Potts model, the percolation process involved is
obtained through the mapping onto the random graph model.
These algorithms are based on the
Fortuin-Kasteleyn representation~\cite{FortuinKasteleyn69} where
bond variables are introduced.
In the Swendsen-Wang algorithm \cite{SwendsenWang87}, a 
cluster update sweep consists of three steps: 
depending on the nearest neighbour 
exchange interactions, assign 
values to the bond variables, 
then identify clusters of spins connected by active
bonds, and eventually assign a random value to 
all the spins in a given cluster. 
The Wolff algorithm~\cite{Wolff89} is a simpler 
variant in which only a single cluster
is flipped at a time. A spin is randomly chosen, 
then the cluster connected with
this spin is constructed and all the spins in the cluster are updated.

Both algorithms considerably improve the efficiency close to the critical
point and their performances are comparable in two dimensions, so in principle
one can equally choose either one of them. Nevertheless, 
when one uses particular boundary
conditions, with fixed spins along some 
surface for example, the Wolff algorithm
is less efficient, since close to criticality the unique cluster will often
reach the boundary and no update is made in this case.

\subsubsection{Definition of the physical quantities\label{SubsecQty}}
For each disorder strength, many samples ($N_{\rm rdm}$) from the same 
probability distribution are studied at a given temperature. Each sample,
initialised from the low-temperature phase, is thermalized during $N_{\rm th}$ 
Monte Carlo sweeps and the physical quantities are then measured during 
$N_{\rm MC}$  sweeps. Different quantities, averaged over the MC sweeps
denoted by $\langle\dots\rangle$, are conserved for all the samples. 
Many physical  quantities can be measured:

\begin{enumerate}
\item[i)]  The order parameter density follows from the standard
definition for the Potts model:
$$        M=\langle\sigma\rangle,\quad \sigma=\frac{q\rho_{\rm max}-1}{q-1},
        $$
where $\rho_{\rm max}$ is the fraction of spins in the majority orientation
$$        \rho_{\rm max}={\rm Max}_\alpha(\rho_\alpha),\quad 
        \rho_\alpha=\frac{1}{L^2}\sum_{j}\delta_{\sigma_j,\alpha}.
  $$
Thermal average is understood in the notation $M$. To obtain the local order
parameter $\langle\sigma(i)\rangle$ at site 
$i$, it is counted 1 when the spin at
site $i$ is in the majority state and 0 otherwise.

\item[ii)]  The susceptibility is given by fluctuation-dissipation theorem
$$        k_BT\chi=\langle\sigma^2\rangle-\langle\sigma\rangle^2.
  $$
 
\item[iii)] Energy density:
$$        E=\langle\varepsilon\rangle,\quad\varepsilon=\frac{1}{2L^2}
        \sum_{(i,j)}K_{ij}
        \delta_{\sigma_i,\sigma_j}.
  $$
\item[iv)] Specific heat:
$$        C/k_B=\langle\varepsilon^2\rangle-\langle\varepsilon\rangle^2.
  $$
\item[v)] Energy Binder cumulant:
$$        U_E=1-\frac{\langle\varepsilon^4\rangle}{3\langle
        \varepsilon^2\rangle^2}.
  $$
\item[iv)] Correlation functions: the connected
spin spin correlation function $G_\sigma(i,j)=\langle\sigma(i)\sigma(j)\rangle
-\langle\sigma^2\rangle$ at criticality is obtained by the estimator  of the
paramagnetic phase,
$$\frac{q\langle\delta_{
        \sigma_i\sigma_{j}}\rangle-1}{q-1},
$$
given by the probability
that spins at sites $i$ and $j$ belong to the same finite cluster.
\end{enumerate}
All these quantities are then averaged over the disorder realisations
$$\overline{\langle...\rangle}=\int \langle...\rangle {\cal P}[\langle...\rangle]{\rm d}\langle...\rangle.$$

\subsection{Transfer matrix technique}
The disordered Potts model can be studied using the 
transfer matrix method introduced by Bl\"ote and 
Nightingale~\cite{BloteNightingale82}, which takes
advantage of the Fortuin-Kasteleyn representation~\cite{FortuinKasteleyn69}
in terms of graphs of the
partition function of the Potts model in order 
to reduce the dimension of the Hilbert 
space~\footnote{A refined algorithm based on a loop representation of the partition function was
proposed in Ref.~\cite{DotsenkoJacobsenLewisPicco99}.}. 
In the Fortuin-Kasteleyn representation, the partition function (with no magnetic
field) is
$$        Z={\rm Tr}\prod_{(i,j)}
        (1+\delta_{\sigma_i,\sigma_j} u_{ij}),
$$ 
where $u_{ij}={\rm e}^{K_{ij}}-1$, is expanded as a
sum over all the possible graphs ${\cal G}$ (with $s$ sites and $l({\cal G})$ loops) 
leading to the random cluster model:
$$        Z=q^s\sum_{\cal G} q^{l({\cal G})}\prod_{(i,j)/b_{ij}=1}
        \left(\frac{u_{ij}}{q}\right),
$$
$b_{ij}\in\{0;1\}$ being the bond variables.
Bl\"ote and Nightingale suggested to introduce a set of connectivity states
which contain the information about which sites on a given row
belong to the same cluster when they are interconnected through a part of
the lattice previously built. A unique connectivity label
$\eta_{i}=\eta$ is attributed to all the sites $i$ of such a cluster. 
In the connectivity space, $|Z(m)\rangle$ is a vector
whose components are given by the partial partition function 
$Z(m,\{\eta_i\}_m)$
of a strip of length $m$ whose connectivity on the last row is given by
$\{\eta_i\}_m$. The connectivity transfer matrix is then defined
according to $|Z(m+1)\rangle={\bf T}_m|Z(m)\rangle$ and the partition 
function of a strip of
length $m$ becomes $|Z(m)\rangle=\prod_{k=1}^{m-1} {\bf T}_k|Z(1)\rangle$,
where $|Z(1)\rangle$ is the statistics of uncorrelated spins.
\begin{figure}[ht]
  \vskip.3mm\epsfxsize=8cm
  \centerline{\epsfbox{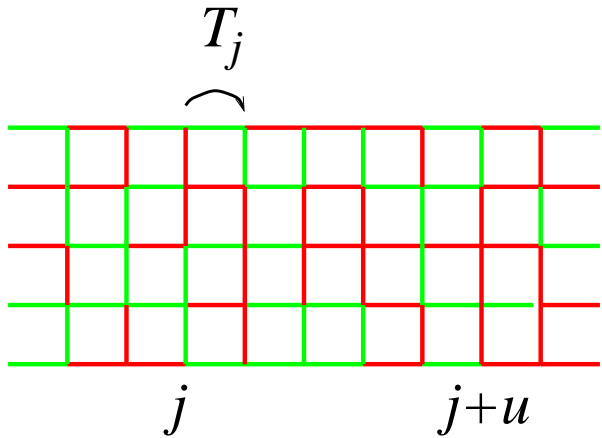}}
  \vskip 0truecm
  \caption{Transfer matrix along the disordered strip.} 
  \label{figruban} 
\end{figure}

The following physical  quantities are measured:
\begin{enumerate}
\item[i)]  The quenched free 
energy density is given (up to a $k_BT$ factor) 
by the  Lyapunov exponent of the
product of an infinite number of transfer matrices
${\bf T}_k$~\cite{Furstenberg63}
\begin{equation}
        \overline{f_L}=-L^{-1}\Lambda_0(L).
\label{eq-F-ave}
\end{equation}  
\begin{equation}
        \Lambda_0(L)=\lim_{m\to\infty}\frac{1}{m}
        \ln\left|\!\left|\left(\prod_{k=1}^m 
        {\bf T}_k\right)
        \mid\! v_0\rangle\right|\!\right|,
\label{eq-Furst}
\end{equation}  
where $\mid\! v_0\rangle$ is a unit initial vector. 

For a pure system, the central charge $c$ is defined as the universal 
coefficient in the lowest-order correction to scaling of
the free energy density ${f_L}$ of a strip of width $L$:
        \begin{equation}
            {f_L}=f_\infty-{\pi c\over 6L^2}
                +\order\left({1\over L^4}\right), 
        \end{equation}  
where the regular contribution is
$        f_\infty=\lim_{L\rightarrow +\infty} {f_L}$.
For a disordered system, $c$ is defined in the same way from the finite-size
behaviour of the quenched average free energy density $\overline{f_L}$, and
numerically, since the strip widths available are small, we can only expect
to measure effective central charges which depend on the disorder strength,
$c_{\rm eff}(g)$, and which would converge towards the true value $c$ in the
thermodynamic limit~\cite{JacobsenPicco00,DotsenkoJacobsenLewisPicco99}.
        \begin{equation}
            \overline{f_L}=f_\infty-{\pi c_{\rm eff}\over 6L^2}
                +a_4L^{-4}. 
        \end{equation}

\item[ii)] 
The spin-spin correlation functions in the time-direction ($u$)  
of the strip are
calculated using an extension of the Hilbert space that allows to keep track
of the connectivity with a given spin.
For a specific disorder realisation, the spin-spin correlation function
along the strip 
\begin{equation}
        G_{\sigma}(u)=\frac{q\langle\delta_{
        \sigma_j\sigma_{j+u}}\rangle-1}{q-1},
        \label{eq-Gu}
\end{equation}  
 is given by the probability that the spins along some row,  
at columns $j$
and $j+u$, are in the same state and is expressed, in the absence of long-range order,
in terms of a product
of non-commuting transfer matrices: 
\begin{equation}
        \langle\delta_{\sigma_j\sigma_{j+u}}\rangle\sim
        \langle \Lambda_0\!\mid{\bf g}_j
        \left(\prod_{k=j}^{j+u-1}
        {\bf T}'_k\right){\bf d}_{j+u}\mid\! \Lambda_0\rangle ,
\label{eq-corr}
\end{equation}
where $\mid\! \Lambda_0\rangle$ is the ground state eigenvector and ${\bf T}_k'$ is the 
transfer matrix in the extended Hilbert space. The operators 
${\bf g}_j$ and ${\bf d}_{j+u}$ realise the mapping between the two connectivity
spaces.
The correlations were computed on strips of varying widths  and then averaged over
many disorder realisations.
\end{enumerate}

\section{Analysis of numerical data in $2D$\label{sec:four}}
\subsection{Location of the random fixed point}
The transition line between ordered and disordered phases in the phase diagram
starts at some point corresponding to the pure system and ends at another
point where the critical properties are governed by the percolation 
universality class. Somewhere between, the random fixed point governs the
critical behaviour of quenched randomness. Although this random fixed 
point is attractive, its precise location is an important preliminary step.
Indeed, if the assumption of the existence of a unique stable  
random fixed point holds, one
expects that the critical behaviour is asymptotically  the same as the 
system is moved 
along the transition
line. 
However, in finite systems, one generically has to deal with
strong crossover effects due to the competition between the disordered
fixed point and the pure  and percolation fixed points, or to corrections 
to scaling linked to the appearance of irrelevant scaling variables. 
These latter
effects are generally important in random systems and the
corresponding corrections to scaling can be
substantially reduced when one measures the critical exponents
in the regime of the random fixed point,  expected to be reached 
at the vicinity of the maximum of the effective central charge.

\begin{figure}[ht]
  \vskip.3mm\epsfxsize=12cm
  \centerline{\epsfbox{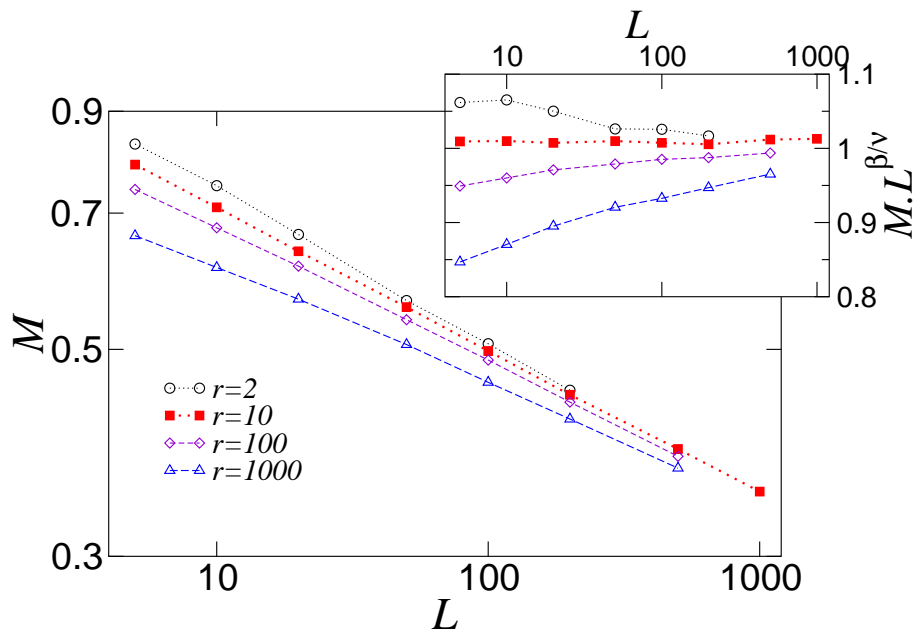}}
  \vskip 0truecm
  \caption{Finite-size scaling behaviour of the magnetisation ($q=8$,
    binary disorder) for
    different disorder amplitudes $r$ (binary distribution). 
    The corrections to scaling are
    smaller close to $r=10$ (taken from Picco~\cite{Picco98}).} 
  \label{figCrossover} 
\end{figure}

\begin{figure}[h]
        \vskip.3mm\epsfysize=9cm
        \begin{center}
        \epsfbox{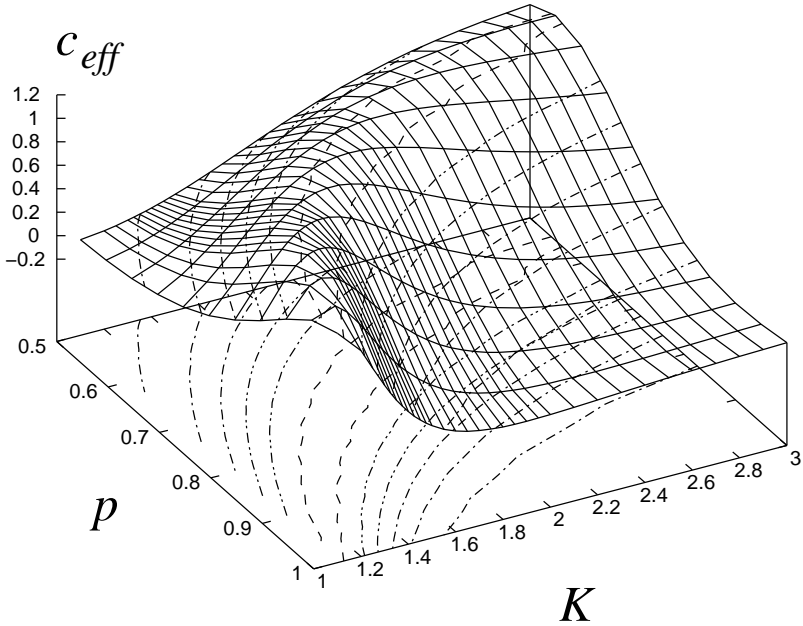}
        \end{center}\vskip 0cm
        \caption{Dependence of the effective central charge $c_{\rm eff}$ 
        with the exchange coupling $K$ and the bond probability $p$ for 
        the 4-state Potts model. The maximum gives the location of the 
        transition line and the absolute maximum corresponds to the
        optimal disorder strength.}
        \label{Figbilan-c_eff}
\end{figure}

Let us consider the finite-size behaviour of an observable $Q$
measured at a deviation $t=K-K_c(g)$ from the critical point on some
system of characteristic size $L$, in the presence of disorder whose strength 
is measured by an amplitude $g$ (ratio $r$ between strong and weak interactions,
probability $p$ of non vanishing bond,\dots). The variables $t$ and $L^{-1}$ 
play the role of relevant
scaling fields (with positive RG eigenvalues $y_t=1/\nu$
and $y_L=1$ respectively), while close to the fixed point, disorder is supposed
to be related to some irrelevant scaling variable with eigenvalue
$y_g=-\omega<0$. At the fixed point there is no need that the irrelevant
scaling field vanishes, so that one can write $g^*$ the corresponding
disorder strength at the fixed point and the observable $Q$ obeys the 
following homogeneity assumption
in the scaling region 
$$        Q(t,L^{-1},g)=L^{-x_Q}f(L^{1/\nu}t,L^{-\omega}(g-g^*)).
$$
An expansion of the last variable (keeping the leading term only) 
along the critical
line (i.e. varying $g$ at $K_c(g)$) gives
$$        Q(0,L^{-1},g)=\Gamma_QL^{-x_Q}(1+\Gamma_Q^{(2)}(g-g^*)L^{-\omega}+
        \dots),
$$
where the $\Gamma$'s are non-universal critical amplitudes.
It is thus possible to fix $g=g^*$ in order to minimise the
corrections to scaling (see figure~\ref{figCrossover}).

\begin{figure}[h]
        \vskip.3mm\epsfxsize=10cm
        \begin{center}
        \epsfbox{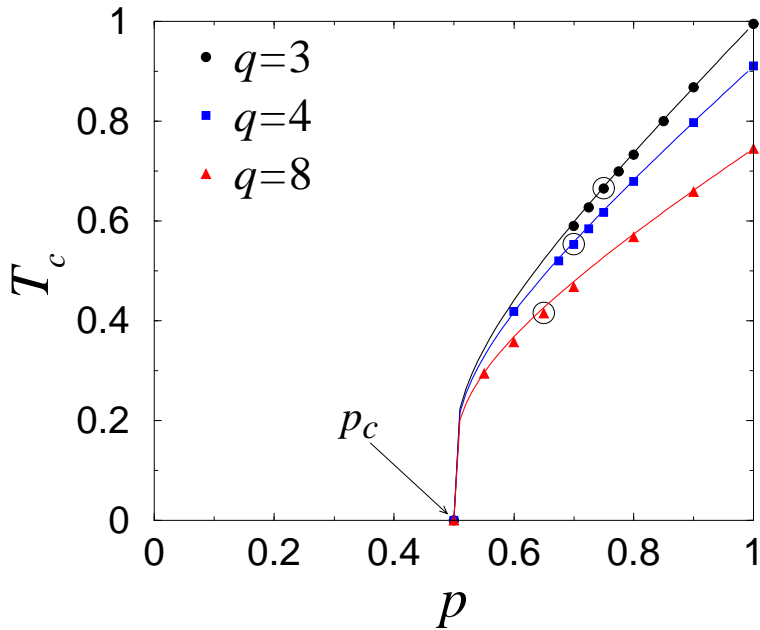}
        \end{center}\vskip 0cm
        \caption{Phase diagram obtained 
          from the condition of a
          maximum of the central charge for the dilute Potts model 
          ($q=3$, 4, and 8). The
        optimal disorder strength (here $p^*$) is denoted by a larger
        circle and the full lines correspond to the single-bond effective
      medium approximation~\cite{Turban80a,Turban80b,GuilminTurban80}.}
        \label{FigPhaseDiagPotts348}
\end{figure}

From a finite-size scaling analysis (see section~\ref{Sec:FSS}), 
the optimal disorder strength $g^*$ is reached when 
a given quantity seems to be well fitted by a simple power-law (i.e.
no bending in the the log-log plot).
In the strip geometry, this value $g^*$ is
found to coincide with the location of the maximum of the central
charge along the critical line~\cite{JacobsenCardy98,ChatelainBerche99},
as a consequence of Zamolodchikov's $c-$theorem~\cite{Zamolodchikov}
(see figure~\ref{Figbilan-c_eff}).

In the literature, different types of disorder distributions have been 
considered. In the most studied case, when self-duality holds, the
exact transition curve is known and the optimal disorder strength
$g^*$ has to be found while moving one parameter only, which simplifies
the task. If one defines the variables $y_{ij}$ by
${\rm e}^{y_{ij}}=\frac{1}{\sqrt q}({\rm e}^{K_{ij}}-1)$, the duality
relation can be written~\cite{Wu82}
$$
{\rm e}^{y^*_{ij}}=\frac{{\rm e}^{K^*_{ij}}-1}{\sqrt q}
=\frac{\sqrt q}{{\rm e}^{K_{ij}}-1}={\rm e}^{-y_{ij}}$$
and self-duality is satisfied when the probability of each coupling
equals the probability of its dual coupling,
\be {\cal P}(y_{ij}){\rm d}y_{ij}={\cal P}(y_{ij}^*){\rm d}y_{ij}^*,\ee
that is if the distribution ${\cal P}(y_{ij})$ is even.
In the case of the bimodal distribution, 
the self-duality point
\begin{equation}
        [\exp(K_c(r))-1][\exp(rK_c(r))-1]=q\;,
\label{duality}
\end{equation}
corresponds to the critical point of the model if only one phase
transition takes place in the system as rigorously shown in
Ref.~\cite{ChayesShtengel98}. 

In the case of dilution, self-duality does not work and the transition line
has to be found numerically. The condition of a maximum of the central charge
is again used as illustrated in figures~\ref{Figbilan-c_eff} and
\ref{FigPhaseDiagPotts348}. The result is in fair agreement with the
effective medium approximation~\cite{Turban80a,Turban80b,GuilminTurban80}.

\subsection{Temperature dependence\label{Sec:T}}
According to their definition, the critical exponents can be obtained
from a temperature-dependence study. Using for
example the case of the order parameter in the low temperature phase,
one can write
$$M(t)=B_-|t|^\beta (1+\dots),\quad t=K_c-K<0.$$
\vskip5mm
\begin{figure}[h]
  \vskip.3mm\epsfxsize=10cm
  \centerline{\epsfbox{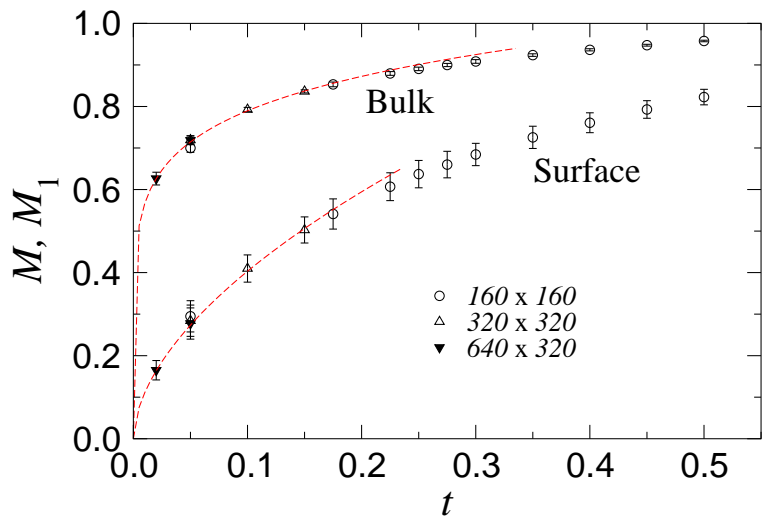}}
  \vskip 0truecm
  \caption{Temperature dependence of the bulk and boundary magnetisation 
    ($q=8$, $r=10$, binary disorder).} 
  \label{figM-de-t}
\end{figure}
The dots indicate the correction terms whose importance depend on the size 
of the system and on the distance from the transition temperature.
Technically, one uses the definition of an effective temperature-dependent
exponent, as illustrated in figures~\ref{figM-de-t} and \ref{figBeta-de-t},
$$\beta_{\rm eff}(t)=\frac{{\rm d}\ln M(t)}{{\rm d}\ln t},\quad
\beta=\lim_{t\to 0}\beta_{\rm eff}(t).$$
 The
precise value of the disorder strength $g$ of course also influences the value
of $\beta_{\rm eff}(t)$ (playing a role in the corrections to scaling
as mentioned above) but asymptotically the limit $t\to 0$ should be
independent on $g$, since there is only one fixed point governing the
disordered system.

\begin{figure}
  \vskip.3mm\epsfxsize=10cm
  \centerline{\epsfbox{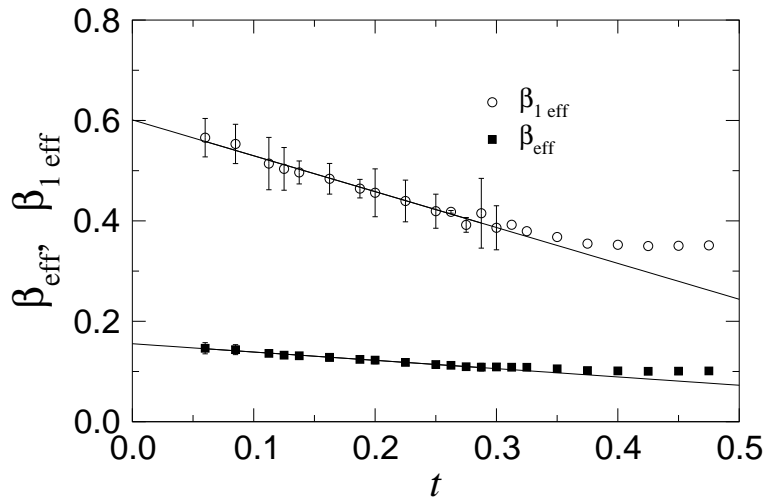}}
  \vskip 0truecm
  \caption{Temperature dependence of the exponents associated to the
    bulk and boundary  magnetisation 
    ($q=8$, $r=10$, binary disorder).} 
  \label{figBeta-de-t} 
\end{figure}

\subsection{Finite-size scaling\label{Sec:FSS}}
\begin{figure}[h]
  \vskip.3mm\epsfxsize=10cm
  \centerline{\epsfbox{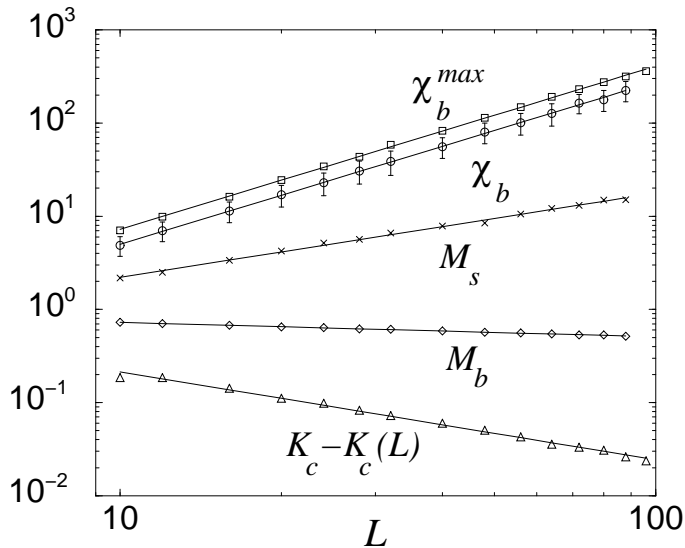}}
  \vskip 0truecm
  \caption{Finite-size scaling analysis of magnetisation, susceptibility
    and effective transition temperature shift for the critical self-dual
    binary disordered $8-$state Potts model.} 
  \label{figFSS} 
\end{figure}

One of the simplest method to extract critical exponents (once the critical
temperature is known) is probably standard Finite-Size Scaling. On a finite
system, the physical quantities cannot exhibit any singularity. They can
be written as a singular term corrected by some scaling function
which depends on the characteristic sizes of the problem, the correlation
length $\xi$ and the size of the system $L$. In the case of the
order parameter density for example we get 
$m_L(T)=|K-K_c|^\beta f(L/\xi)$. The function $f(x)$ of course depends on
the geometry, but at the critical point $K_c$, the following behaviour 
is obtained:
\be 
M_L(K_c)\simunder_{L\to\infty} L^{-\beta / \nu}.\ee  
Here, the ratio
$\beta/\nu$ is precisely the magnetic scaling dimension $x_\sigma$. An example
is shown in figure~\ref{figFSS} for the $8-$state Potts model. From
the slopes of the curves, the values  $\gamma/\nu=1.686(17)$,
$\beta/\nu=0.152(4)$ and $\nu=1.005(30)$ can be obtained~\cite{Chatelain00}. 
The results here are interesting
as reference values that we shall compare with more sophisticated techniques
later.

\subsection{Short-time dynamics scaling}
It is commonly believed that universality can be found only in 
equilibrium stage of long-time regime in numerical simulations. 
For a magnetic system far from criticality, e.g. in the high temperature
phase, suddenly quenched to the critical temperature, a universal 
dynamic scaling behaviour
emerges already within the short-time regime, according to a simple
generalisation of the homogeneity 
assumption for the order 
parameter~\cite{YingHarada00,PanEtal01,YingEtal01,DeroulersYoung02},
\be
M(t,\tau,L,M_0)=b^{-\beta/\nu}M(b^{1/\nu}t,b^{-z}\tau,b^{-1}L,b^{x_0}M_0),
\ee
where $z$ is the dynamic exponent (dependent on the choice of algorithm),
$t=|K-K_c|$ is the deviation from the critical point, $M_0$ is the initial
magnetisation with the associated scaling dimension $x_0$, and $\tau$
is the time (measured in MC sweeps).
In the thermodynamic limit, $L\to\infty$, and at criticality, $t=0$, the 
expected evolution is given by
$M(\tau,M_0)=\tau^{-\beta/\nu z}f(M_0\tau^{-x_0/z})$ and allows the
computation of the critical exponents.
The main interest of short-time dynamics scaling is that it is not
affected by critical slowing down, since only early time
stages of the simulation are involved.

\subsection{Conformal mappings}
\begin{figure}[h]
        \vskip.3mm\epsfxsize=6cm
        \mbox{\epsfbox{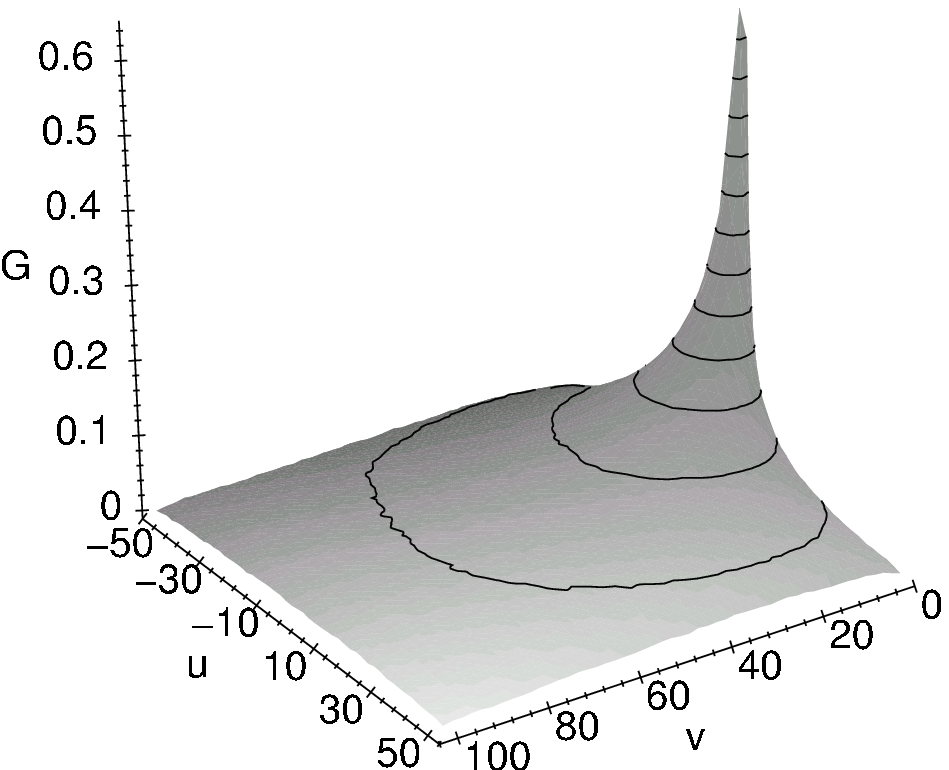}\hfill\qquad}
        \vskip -4.5cm
        \vskip.3mm\epsfxsize=4cm
        \mbox{\qquad\qquad\qquad\qquad
        \qquad\qquad\qquad\qquad\qquad\epsfbox{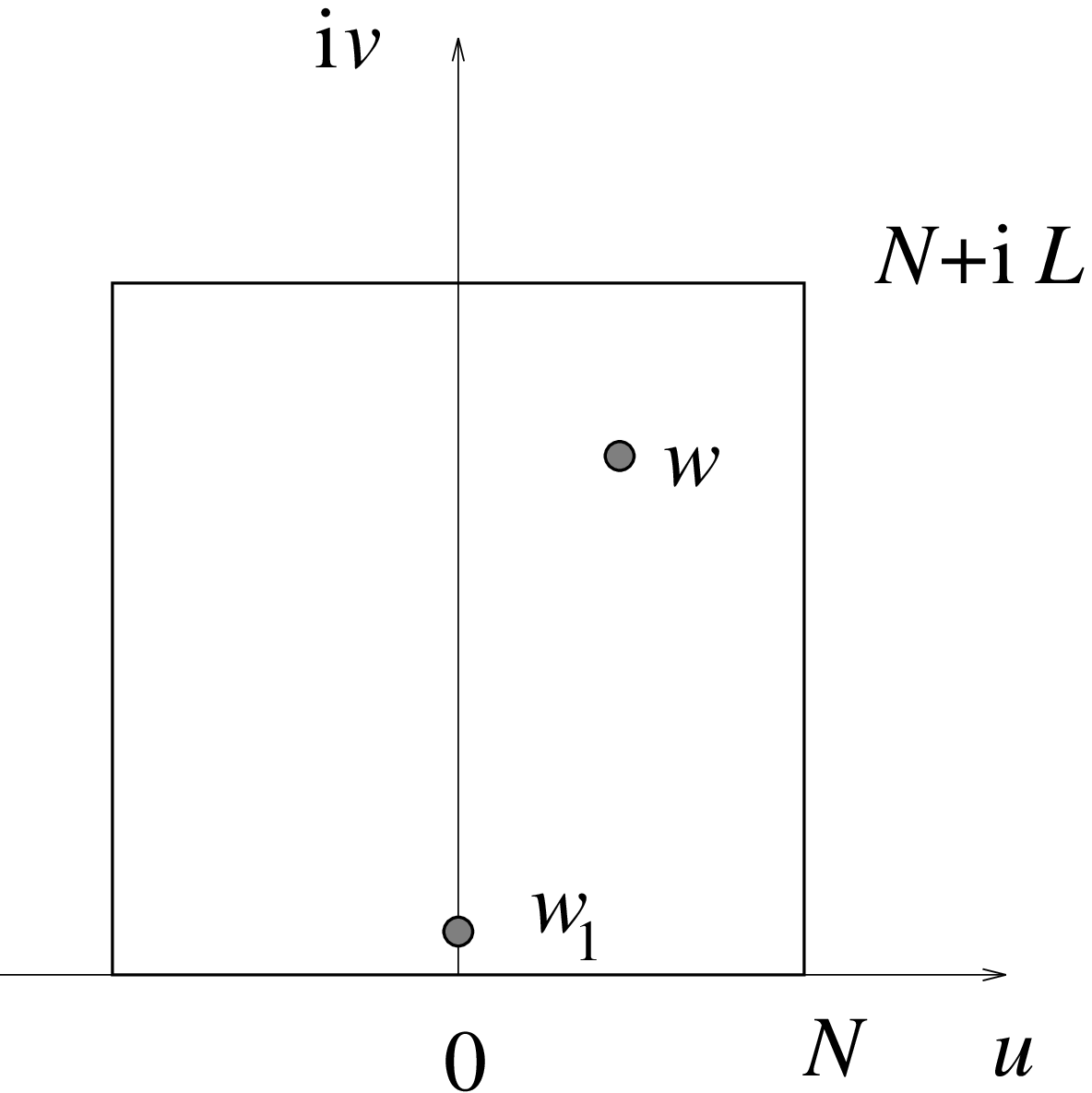}}
        \vskip 0.3cm\caption{Monte Carlo simulations of the 2$d$ 
          Ising model inside 
        a square of
        $101\times 101$ lattice sites
        ($10^6$ MCS/spin, Swendsen-Wang cluster algorithm). The 
        figure shows the correlation function between a point
        close to the surface ($w_1={i}$) and all 
        other points $w$ in the 
        square. The sketch on the right specifies the notations.}
        \label{fig: 1}  
\end{figure}

Monte Carlo simulations of two-dimensional spin systems are 
generally performed on systems of
square shape while transfer matrix computations are done in strip
geometries.  
In the following, we consider such a  system of size
$2N\times L$, and call $u$ and $v$ the corresponding directions
(figure~\ref{fig: 1}). 
The order parameter correlations between a point close to 
the surface, and a
point in the bulk of the system should, in principle, lead to both surface and
bulk critical exponents. Practically, it
 is not of great help for the accurate determination of critical exponents,
since 

i) strong surface effects (shape effects) occur which modify the 
large distance power-law 
behaviour,

ii) the universal scaling function entering the correlation 
function is likely to display a crossover before its asymptotic 
regime is reached (system-dependent effect).

One can proceed as follows: systems of increasing sizes 
 are successively
considered, and the correlations are computed along $u$- 
(parallel to a square edge considered as the free surface) and $v$-axis 
(perpendicular
to this edge). The order parameter correlation 
function for example is supposed to obey a scaling
form which reproduces the expected power-law behaviour in the thermodynamic 
limit:
\begin{equation}
  G_\perp(v)=\frac{1}{v^{x_\sigma+x_\sigma^1}}
  f_{\ \!_{\opensquare}}\left(\frac{v}{L}\right),\qquad
  G_\parallel(u)=\frac{1}{ u^{2x_\sigma^1}}
  f'_{\ \!_{\opensquare}}\left(\frac{u}{N}\right),
  \label{eq: scalG}
\end{equation}
where $x_\sigma$ and $x_\sigma^1$ are, respectively, the bulk and surface 
order parameter scaling dimensions. The scaling functions 
$f_{\ \!_{\opensquare}}$ depends on the geometry, but have to satisfy  
asymptotic expansions including corrections to scaling e.g.
$f_{\ \!_{\opensquare}}\left(\frac{v}{L}\right)\sim 1+{\rm const}
\left(\frac{v}{L}\right)^\epsilon+\dots$ in the boundary region $v\to L$.


\begin{figure}[ht] 
        \vskip.3mm\epsfxsize=11cm
        \begin{center}\vglue -0cm
        \epsfbox{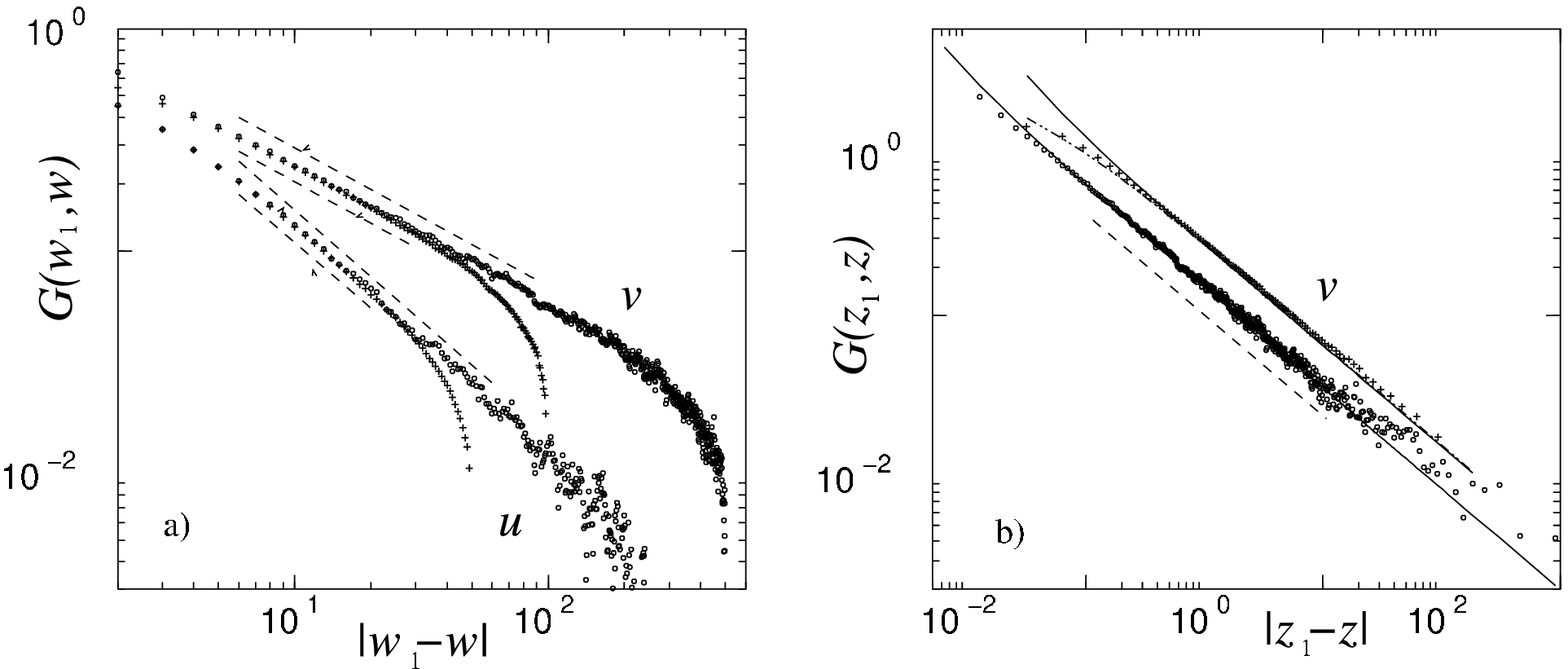}
        \end{center}
        \vskip -8mm\caption{Monte Carlo simulations in the case of the 
        two-dimensional
        Ising model. \hfill\break 
        a) Log-log plot of the order parameter correlation
        function perpendicular to the surface ($v$-direction) and parallel to
        this surface ($u$-direction) for system sizes $101\times 101$ ($+$,
        $5.10^6$ MCS/spin) and
        $501\times 501$ ($\circ$, $2.10^5$ MCS/spin). 
        The correlations are computed between a point
        $w_1={i}$ close to the surface
        and points $w={i}v$, and $w=u+{i}$ respectively.
        The corresponding estimations for $x_\sigma+x_\sigma^1$ and 
        $2x_\sigma^1$ are written
        in the figure
        and the fit has been done in the range 
        indicated by a dashed line. The figures 
        correspond to power-law fits for the size $501^2$.\break 
        b) Conformal rescaling of the perpendicular correlations 
        for the two sizes (see below).
        The rescaled correlation function 
        exhibits a true power-law in the whole range of variation 
        of the new variable, and the value for $x_\sigma+x_\sigma^1$ is 
        improved. The solid lines correspond to the theoretical asymptotic form
        while the dotted line is the exact expression in the continuum limit.}
        \label{fig: 2}  
\end{figure}

Equations~(\ref{eq: scalG}) are not very 
useful for the determination of critical
exponents, since at least five unknown quantities appear, and the correction 
terms in $f_{\ \!_{\opensquare}}(x)$ may have a large amplitude, 
resulting from the significance
of finite-size corrections. Nevertheless,  if the power-law fit is limited to 
the linear
regions on a log-log scale
(almost one decade in  figure~\ref{fig: 2}a), one gets 
correct estimations 
of the critical exponents, as long as condition ii) is fulfilled.  

\begin{figure}[t]
\vbox{ 
  \vskip 1truecm
  {\par\begingroup
    \parindent=0pt
    \baselineskip=12truept\vskip.3mm\epsfxsize=5truecm
    \ \hfill{\epsfbox{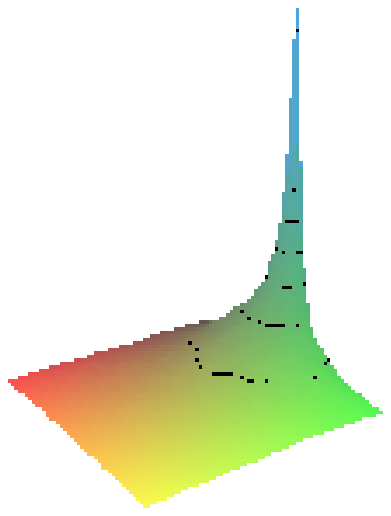}}
    \vskip 0mm
    \par
    \endgroup \par}

  \vskip -7truecm

  {\par\begingroup
    \parindent=0pt
    \baselineskip=12truept\vskip.3mm\epsfxsize=5truecm
    {\epsfbox{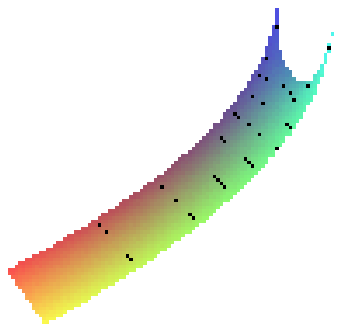}}\hfill
    \vskip 0mm
    \par
    \endgroup \par}

  \vskip-9truecm
  {\par\begingroup
    \parindent=0pt
    \baselineskip=12truept\vskip.3mm\epsfxsize=4truecm
    \centerline{\epsfbox{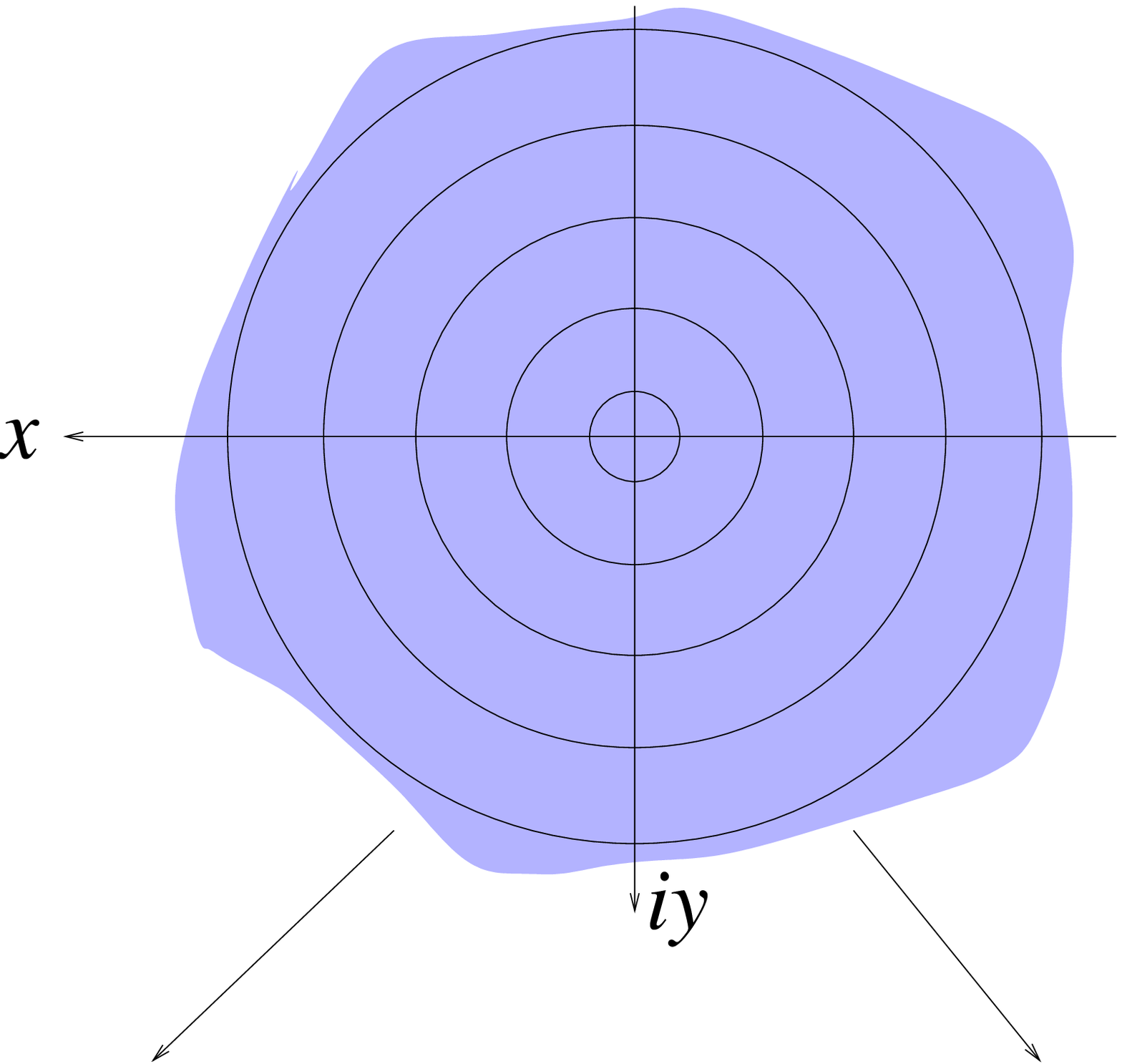}}
    \vskip 0mm
    \par
    \endgroup \par}
  \vskip 4.5truecm
} 
  \caption{Conformal mapping of the infinite plane $z=x+iy$ inside a finite
    square $N\times N$ with free edges or a strip of width $L$ with periodic 
    boundary conditions.} 
  \label{figMapping} 
\end{figure}

At a critical point, scale invariance coupled with rotation and translation 
invariance also implies covariance under local scale transformations, i.e. 
conformal transformations~\cite{Cardy96}.
The conformal transformations are those general
coordinate transformations which preserve the angle
between any two vectors. They leave the metric
invariant up to a scale change. The conformal group
includes the Euclidean group as a subgroup, the
dilatations and the special conformal transformations.
In any dimension these conformal transformations are
also called global conformal transformations because
they map the infinite space onto itself.
Consider a lattice model described in the continuum
limit by a local theory defined  by some action $S$.
Conformal symmetry occurs when a local theory is scale
invariant. In lattice systems, nearest--neighbour
interactions ensure that physics is local and scale
invariance holds at the critical point where a
continuous limit description is allowed.
For any local field (the energy density or the magnetisation for a spin system for example) 
the usual homogeneity assumption under a
homogeneous rescaling ${\bf R}\to b{\bf R}$ 
\be
\langle \phi(0)\phi(b{\bf R})\rangle = b^{-2x_\phi}
\langle \phi(0)\phi({\bf R})\rangle\label{eqCorrScal}\ee
 is extended to local
transformations with a position dependent rescaling
factor.
In two dimensions conformal transformations are realized
 by the analytic functions in the complex plane (the conformal group
is thus infinite-dimensional):
$z\longrightarrow w(z)$ and equation~(\ref{eqCorrScal})
is thus 
generalised to a covariance law of transformation of the ($N-$point) 
correlators under conformal mappings:
\be
\langle \phi(w_1)\phi(w_2)\rangle
=\mid\! w'(z_1)\!\mid^{-x_\phi}
        \mid\! w'(z_2)\!\mid^{-x_\phi} 
\langle \phi(z_1)\phi(z_2)\rangle.
\label{eq-CovG}\ee
Here, $z_1$ and $z_2$ are two points in the original complex plane and
$w_1$ and $w_2$ are the corresponding points in the transformed
complex geometry under the mapping $w(z)$.
This transformation law is very helpful in numerical analysis, since 
simulations or numerical computations are always performed on finite 
systems of particular shape, depending on the technique used. The critical
properties of an infinite system
$ \langle \phi(z_1)\phi(z_2)\rangle\sim |z_1-z_2|^{-2x_\phi}$
can thus be obtained by fitting the numerical
data to the transformed conformal expression 
which usually deviates significantly from
a simple power law (although the algebraic decay at criticality must
be recovered asymptotically in the limit of an infinite system
of course).
The situation is schematically sketched in figure~\ref{figMapping}

Among all the possible mappings, we shall here specify a few cases of
interest:

\begin{description}
\item[i)] Mapping onto a cylinder: the logarithmic transformation
\be w(z)=\frac{L}{2\pi}\ln z=u+iv\ee 
is well known to map the infinite plane
onto a strip of finite width $L$ with periodic boundary conditions and
infinite length. This is the limit $N\to\infty$ in the rectangular geometry
described above. 
Using the algebraic decay of the two-point correlator in
the $z-$plane, one gets on the strip
\be
\langle \phi(0,0)\phi(u,v)\rangle=\left(\frac{2\pi}{L}\right)^{2x_\phi}
\left[2\cosh\left(\frac{2\pi u}{L}\right)
-2\cos\left(\frac{2\pi v}{L}\right)\right]^{-x_\phi}.
\ee
In the long direction of the strip, at large distances it becomes an exponential 
decay 
\be
\langle \phi(0,0)\phi(u,0)\rangle_{\rm pbc}=\left(\frac{2\pi}{L}\right)^{2x_\phi}
\exp\left({-\frac{2\pi u x_\phi}{L}}\right).
\ee

For sufficiently large strip widths, the transverse direction can also
give some interesting results. Using the mapping $w(z)=\frac{L}{\pi}\ln z$,
the half-infinite plane is mapped onto a strip with open boundaries in the 
transverse direction. If the boundary conditions are fixed (for example
using an ordering surface field coupled to the order parameter)
on one edge and free on the opposite edge
(this is the meaning of the notation $+f$ below), 
the transverse profile of the
order parameter density is given by the conformal expression
\be
\langle\sigma(v)\rangle_{+f}={\rm const}\ \!\times
\left[\frac L\pi\sin\left(\frac{\pi v}{L}\right)
\right]^{-x_\sigma}F\left[\cos\left(\frac{\pi v}{2L}\right)
\right].
\ee
The shape of the scaling function $F(x)$ is asymptotically constrained 
by simple scaling, $F(x)\sim x^{x_\sigma^1}$ (here,  
$x_\sigma^1$ is the boundary scaling dimension of the order parameter).

\begin{figure}[h]
  \vskip.3mm\epsfxsize=9cm
  \centerline{\epsfbox{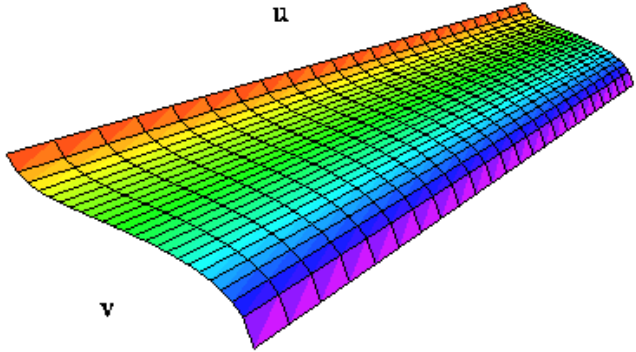}}
  \vskip 0truecm
  \caption{Transverse profiles of the order parameter in an infinitely
    long strip with 
    fixed-free boundary conditions.} 
  \label{figFf3d} 
\end{figure}

\item[ii)] Mapping onto a square: the Schwarz-Christoffel transformation
\be
w(z)={N\over 2{\rm K}}{\rm F}(z,k),
\quad z={\rm sn}\left({2{\rm K}w\over N}\right)\label{eq-SchChr}\ee
maps the half-infinite plane $z=x+iy$ ($0\le y<\infty$) 
inside a square $w=u+iv$ of size
$N\times N$ ($-N/2\le u\le N/2$, $0\le v\le N$)
with free boundary conditions
along the four edges. Here, $F(z,k)$ is the elliptic integral of the 
first kind, ${\rm sn}\ \! (2{\rm K}w/ N)$ 
the Jacobian elliptic sine, ${\rm K}=K(k)$ the
complete elliptic integral of the first kind, and the modulus $k$ is solution
of $K(k)/K(\sqrt{1-k^2})=\frac 12$.

In the semi-infinite geometry, the two-point correlator is fixed up
to an unknown scaling function (apart from some asymptotic limits
implied by scaling). Fixing one point $z_1$ close to the free surface 
($z_1=i$) of
the half-infinite plane, and leaving the second point exploring the
rest of the geometry, $z_2=z$, the following behaviour is expected:
\be\langle \phi(z_1)\phi(z)\rangle_{\hlfplane}
\sim y^{-x_\sigma}\psi(\omega),\ee
where the dependence on $\omega=\frac{y_1y}
  {\mid z_1-z\mid^2}$ of the universal scaling function $\psi$ is constrained
by the special conformal transformation and its asymptotic behaviour,
$\psi(\omega)\sim\omega^{x_\phi^1}$, in the limit $y\gg 1$,
 is implied by scaling. 

Using the mapping~(\ref{eq-SchChr}), the local rescaling factor in 
equation~(\ref{eq-CovG}) is obtained,
$w'(z)=\frac{N}{2{\rm K}}[(1-z^2)(1-k^2z^2)]^{-1/2}$,
and
inside the square the two-point correlation function 
becomes (see Ref.~\cite{ChatelainBerche99})
\be
\langle\phi(w_1)\phi(w)\rangle_{\rm sq.} \sim {\underbrace{
\left(\Im [z].
\left(|1-z^2|.|1-k^2z^2|\right)^{-1/2}
\right)
}_{\kappa(w)}}^{-x_\phi}\psi(\omega),
\ee
with $z(w)$ given by equation~(\ref{eq-SchChr}). 
This expression is correct up to a constant amplitude determined by
$\kappa(w_1)$ which is kept fixed, but the function $\psi(\omega)$ is still
varying with the location of the second point, $w$. 

\begin{figure}[t]
  \vskip-5mm\epsfxsize=9cm
  \centerline{\epsfbox{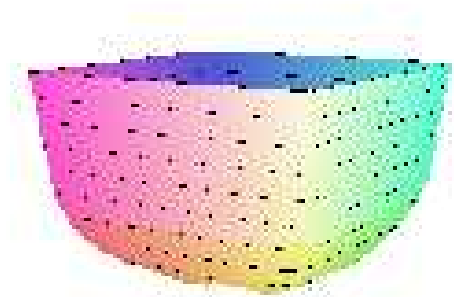}}
  \vskip -3truecm
  \caption{Profiles of the order parameter in a square with 
    fixed boundary conditions.} 
  \label{figProfSq3d} 
\end{figure}

In order to cancel the role of the unknown scaling function,
it is more convenient to work with a density profile in the presence
of ordering surface fields. This is a
one-point correlator whose functional shape 
in the half-infinite geometry is determined by scaling apart
from some amplitude:
\be \langle\sigma(z)\rangle_{\hlfplane} 
= {\rm const\times} y^{-x_\sigma}\ee
and it maps onto
\be
\langle\sigma(w)\rangle_{\rm sq.} =
        {\rm const\times}[\kappa(w)]^{-x_\sigma}
        \ee
where the function $\kappa(w)$ again comes from the mapping.

The case of the mapping of the infinite complex plane inside a square with
periodic boundary conditions was for example used in 
Ref.~\cite{TalapovDotsenko93}.
\end{description}

Other mappings can be convenient. Our choice here was motivated by the 
fact that Monte Carlo simulations are usually performed on samples
of square shapes.
On the other hand, the strip (with free or fixed boundary conditions) 
is the natural geometry generated in transfer matrix calculations.

\subsection{Impact of rare events and non self-averaging}
As we mentioned, the physical properties have to be averaged over many
samples produced with a given probability distribution. One can typically 
encounter two opposite situations, depending on the quantities
under interest. The average order parameter for example is defined as
a sum of quantities affected by randomness,
$$\overline{\langle\sigma\rangle}=\frac{1}{N^2}\overline{\sum_i\langle
\sigma_i\rangle}$$
while the correlation function
$$        \langle\delta_{\sigma_j\sigma_{j+u}}\rangle\simeq
        \langle \Lambda_0\!\mid{\bf g}_j
        \left(\vphantom{\prod}\right.\prod_{k=j}^{j+u-1}
        {\bf T}'_k \left.\vphantom{\prod}\right)
        {\bf d}_{j+u}\mid\! \Lambda_0\rangle$$
essentially depends on a product of non-commuting matrices whose
elements are determined by the disorder distribution. These two types
of quantities definitely exhibit different properties as functions
of the number of samples used to sample the probability distribution.
When computing mean values, it is clear that the accuracy of the results 
for a given number of disorder realizations, does not have the same behaviour
in the case  of sums or of products of random variables. 
Consider as an example the sum and the product of random variables
$\lambda_i$ taken from a binary distribution,
$$\Sigma_\lambda=\sum_{i=1}^n\lambda_i,\quad
\Pi_\lambda=\prod_{i=1}^n\lambda_i,$$
and compute the moments (rescaled by a power $1/p$)
$[\overline{(\Sigma_\lambda)^p}]^{1/p}$
and $[\overline{(\Sigma_\lambda)^p}]^{1/p}$ averaged over $N$ realizations
of the $\lambda_i$'s (here we choose for this example $n=50$, so there are
some $10^{15}$ configurations).
Numerical results are given in the table~\ref{tab3} for some choice of
parameters.

{\small
\begin{table}[ht]
\begin{center}
\begin{tabular}{l|llll|llll}
\hline
&\multicolumn{4}{c}{$\Sigma_\lambda$} & \multicolumn{4}{c}{$\Pi_\lambda$} \\
\hline
$N$ & $p=1$ & $p=5$ & $p=20$ & $p=50$ & $p=1$ & $p=5$ & $p=20$ & $p=50$ \\
\hline
$10^1$ & 0.990 & 1.060 & 1.228 & 1.305 & 1.054 & 1.529 & 2.042 & 2.188 \\  
$10^3$ & 1.014 & 1.049 & 1.156 & 1.298 & 1.060 & 1.292 & 2.143 & 2.630 \\  
$10^5$ & 1.010 & 1.047 & 1.161 & 1.327 & 1.054 & 1.303 & 2.480 & 3.467 \\  
$10^7$ & 1.010 & 1.047 & 1.160 & 1.322 & 1.055 & 1.301 & 2.511 & 3.857 \\  
\hline
\end{tabular}
\end{center}
\caption{Comparison between the moments of a sum and of a product of random
variables distributed according to a bimodal probability distribution, as
a function of the number of realizations.
\label{tab3}}
\end{table}
}

It is particularly clear that the values obtained
from the sum $\Sigma$ converge rapidly
(the variations between results obtained
at different number of realizations correspond to a statistical noise).
This is due to the fact that $\Sigma$ is normally distributed,
while in the case of the product $\Pi$, we note a continuous increase of
the numerical estimate of a given moment as the number of samples 
increases and this effect is
especially pronounced for high moment order $p$. This is due to the fact that
the distribution of the $\Pi$'s is log-normal  
and thus there exist some events which have
a dominant role in the average, but which are so rare that they are not
scanned by a poor statistics which would essentially explore the region of
typical values. In order to be more precise in the distinction between
typical and average value, we rewrite $\ln\Pi$ as 
a sum of random variables, $\ln\Pi_\lambda=\sum_i\ln\lambda_i$, which, 
according to the central limit theorem, has a Gaussian distribution in the 
limit of large number of draws. The typical value corresponds to the maximum
of the probability distribution, $\Pi_{\rm typ}={\rm e}^{\overline{\ln\Pi}}$
where the Gaussian is centered, and clearly differs from the average value
$\overline{\Pi}=\overline{{\rm e}^{\ln\Pi}}$.

\begin{figure}[ht]
  \vskip-10mm\epsfxsize=7.5cm
  \centerline{\epsfbox{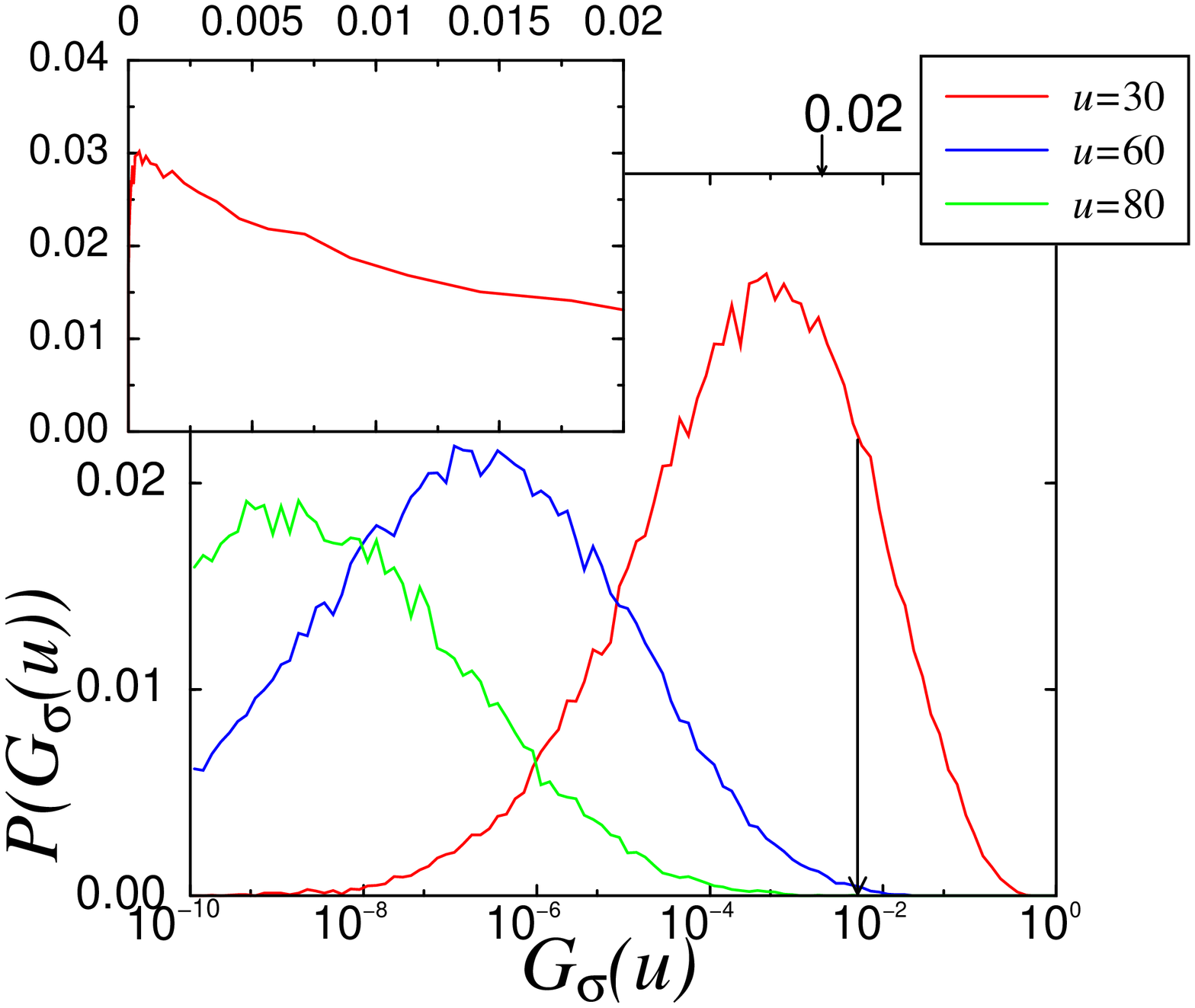}}
  \epsfxsize=7cm
  \centerline{\epsfbox{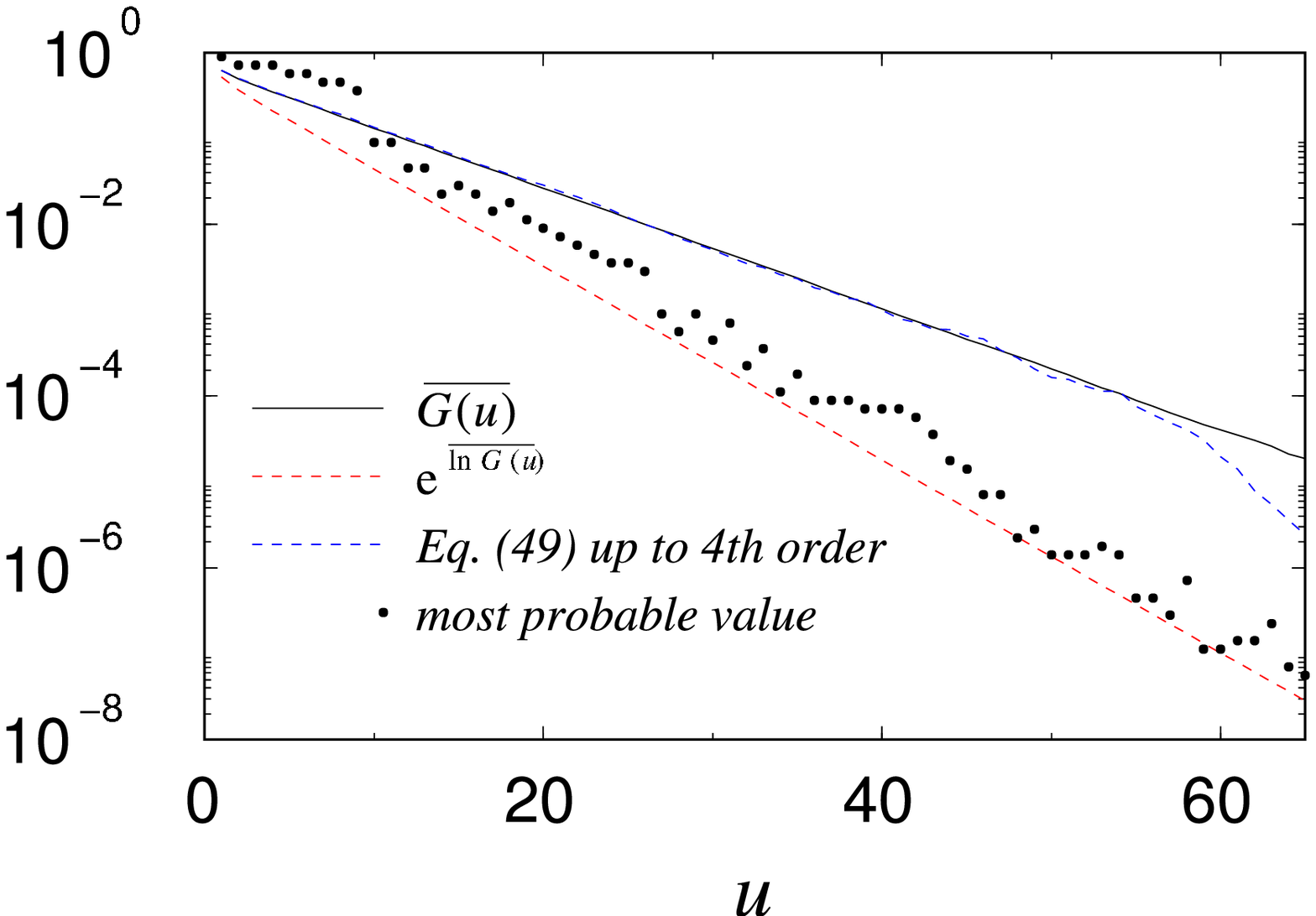}}
  \vskip 0.5truecm
  \caption{Top: Probability distribution of the spin-spin correlation 
    function (8-state Potts model with bimodal disorder). The
    insert shows ${\cal P}(G_\sigma(u))$ (with a very long tail on the right), 
    while it is shown on a logarithmic
    scale in the main frame, where one can notice the shape which is close
    to a log-normal distribution. Bottom: Reconstruction of the correlation function from the moments
    of its logarithm.} 
  \label{figDistG} 
\end{figure}

The same situation occurs when computing the 
spin-spin correlation function 
\be
\overline{\langle\sigma(0)
\sigma(u)\rangle}_{\rm st}=\overline{G_\sigma(u)}.
\ee 
Since it is almost 
log-normal (see figure~\ref{figDistG}), the 
logarithm of $G_\sigma(u)$ is self-averaging and the average
$\overline{\ln G_\sigma(u)}$ 
as well as higher order moments are well behaved. A
 cumulant expansion is thus convenient to reconstruct the average
$\overline{G_\sigma(u)}$ through
\begin{equation}
  \overline{G_\sigma(u)}={\rm e}^{\overline{\ln G_\sigma(u)}+
    \frac 12(\overline{\ln^2 G_\sigma(u)}
    -\overline{\ln G_\sigma(u)}^2)+\dots}.
\end{equation}
This is a test (see figure~\ref{figDistG}) which 
proves that the probability distribution is
sufficiently well scanned with the large numbers of realizations
used in this run.

\newpage
\section{Numerical results and comparison with perturbative expansions
in $2D$\label{sec:five}}
\subsection{Regime $q> 4$}
\subsubsection{Randomness induces a second-order regime}
\begin{figure}[t]
  \vskip.3mm\epsfxsize=8cm
  \centerline{\epsfbox{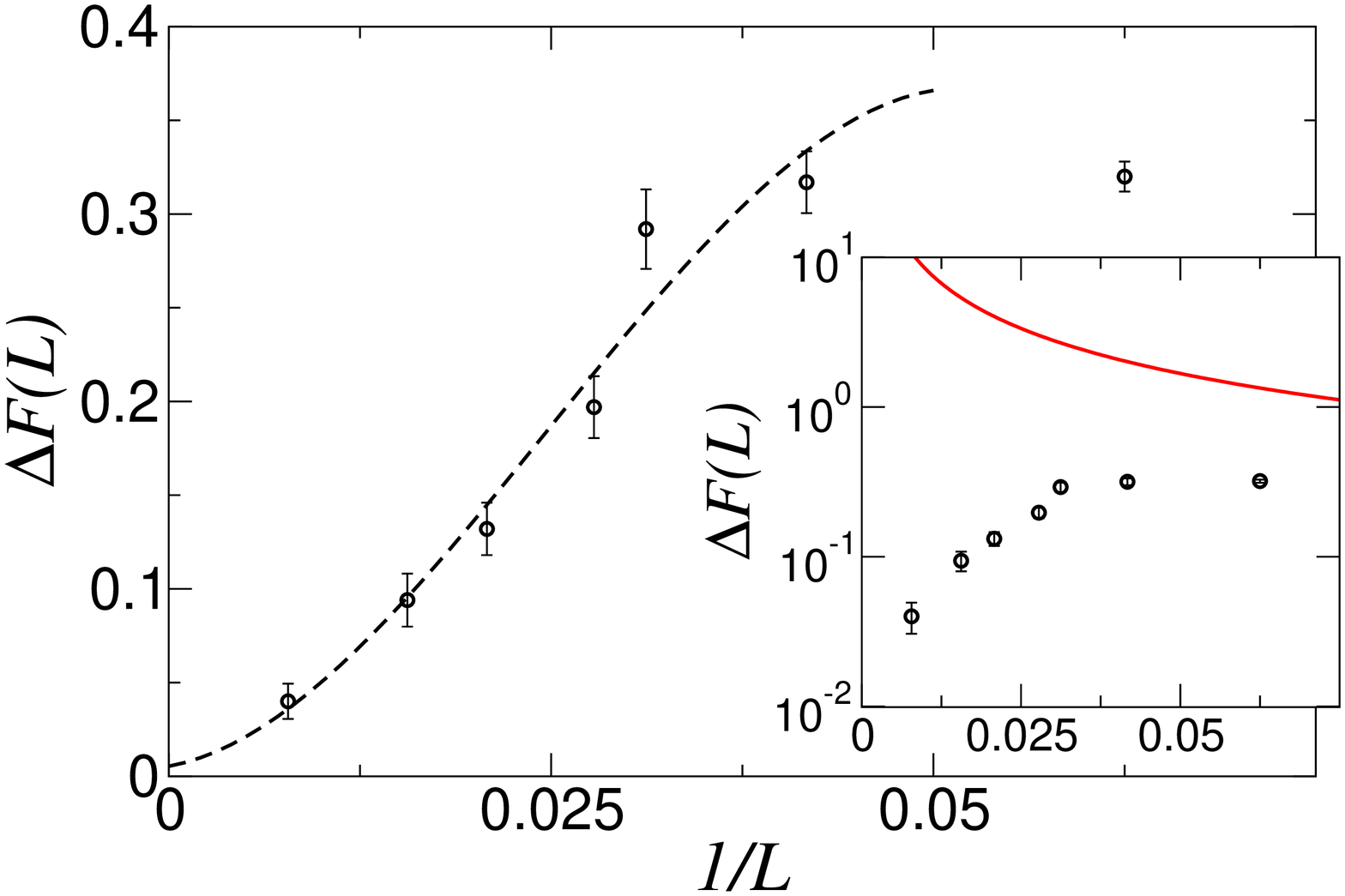}}
 \vskip8mm\epsfxsize=8cm
  \centerline{\epsfbox{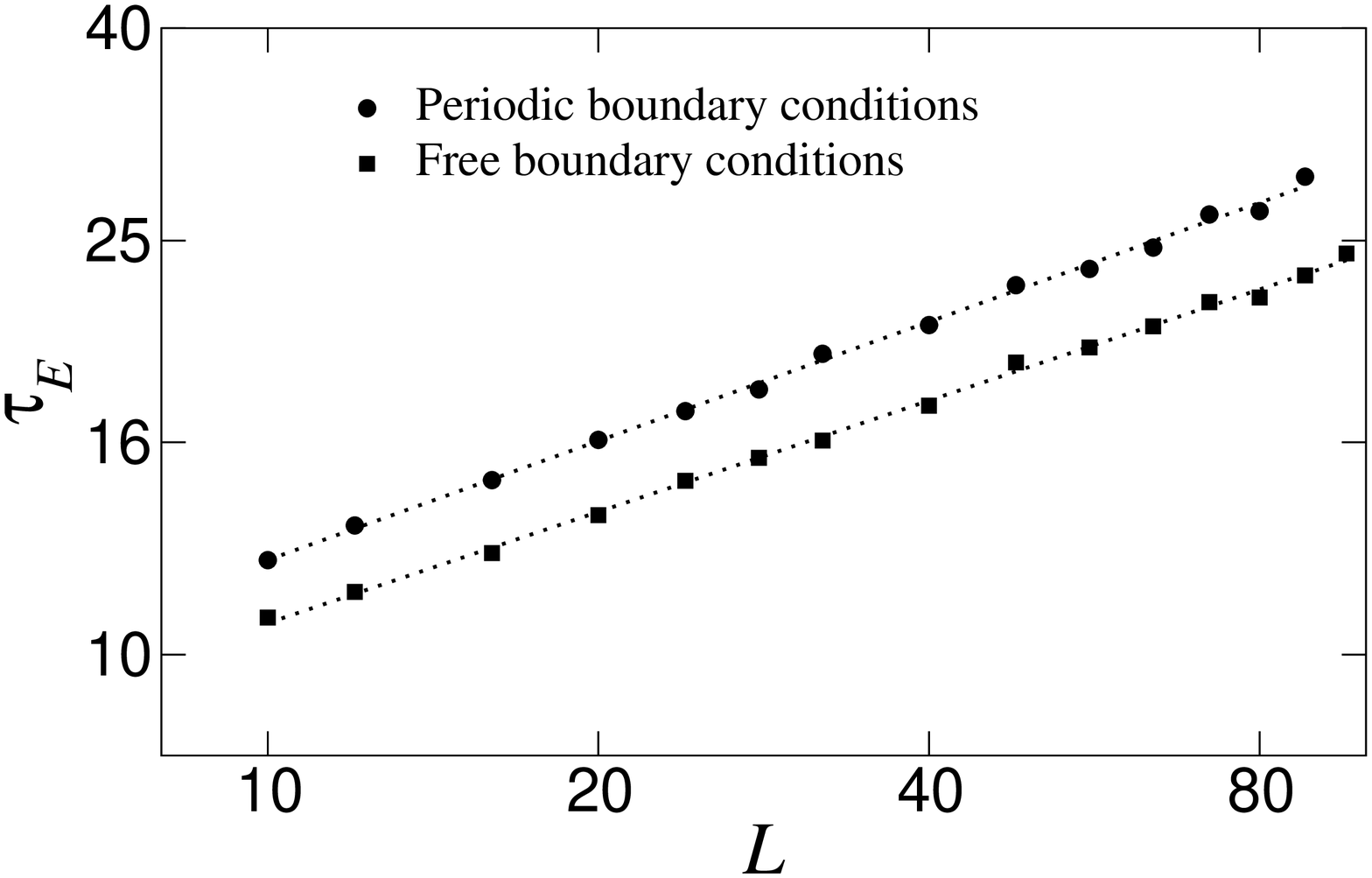}}
  \vskip 0truecm
   \caption{Top: Evolution of the free energy barrier with the size of the system
    for the $8-$state Potts model with binary disorder (at small disorder
    strength). The dotted line in the inset shows the case of the pure
    system (taken from Chen, Ferrenberg and Landau
    Ref.~\cite{ChenFerrenbergLandau95}). The dotted line is a guide
    for the eyes. 
    Bottom: Power-law behaviour of the energy autocorrelation time in the
    $8-$state Potts model with binary disorder.} 
  \label{figZ} 
\end{figure}

Although there is no perturbation result for $q> 4$, we shall start
in the regime of first-order transition of the pure model.
A first step was to prove that even for large numbers of states per spin,
the transition was rounded to become continuous. That was done by
different authors~\cite{ChenFerrenbergLandau95,CardyJacobsen97}.
Chen et al studied the free energy barrier $\Delta F(L)$, defined
from the energy histogram ${\cal P}(E)$ 
in Monte Carlo simulations according to 
${\rm e}^{-\beta\Delta F(L)}=P_{\rm max}/P_{\rm well}$, with $P_{\rm max}$
given by the maximum of ${\cal P}(E)$ 
and $P_{\rm well}$ corresponding to the value at 
the bottom of the well separating the two coexisting phases.
They showed that the energy barrier
$\Delta F(L)=-2\sigma_{\rm o.d.}L^{d-1}$ 
vanishes in the thermodynamic limit (figure~\ref{figZ})
where $\sigma_{\rm o.d.}$ is the order-disorder
interface tension between the two possibly coexisting  phases. 

\begin{figure}[ht]
  \vskip.3mm\epsfxsize=8cm
  \centerline{\epsfbox{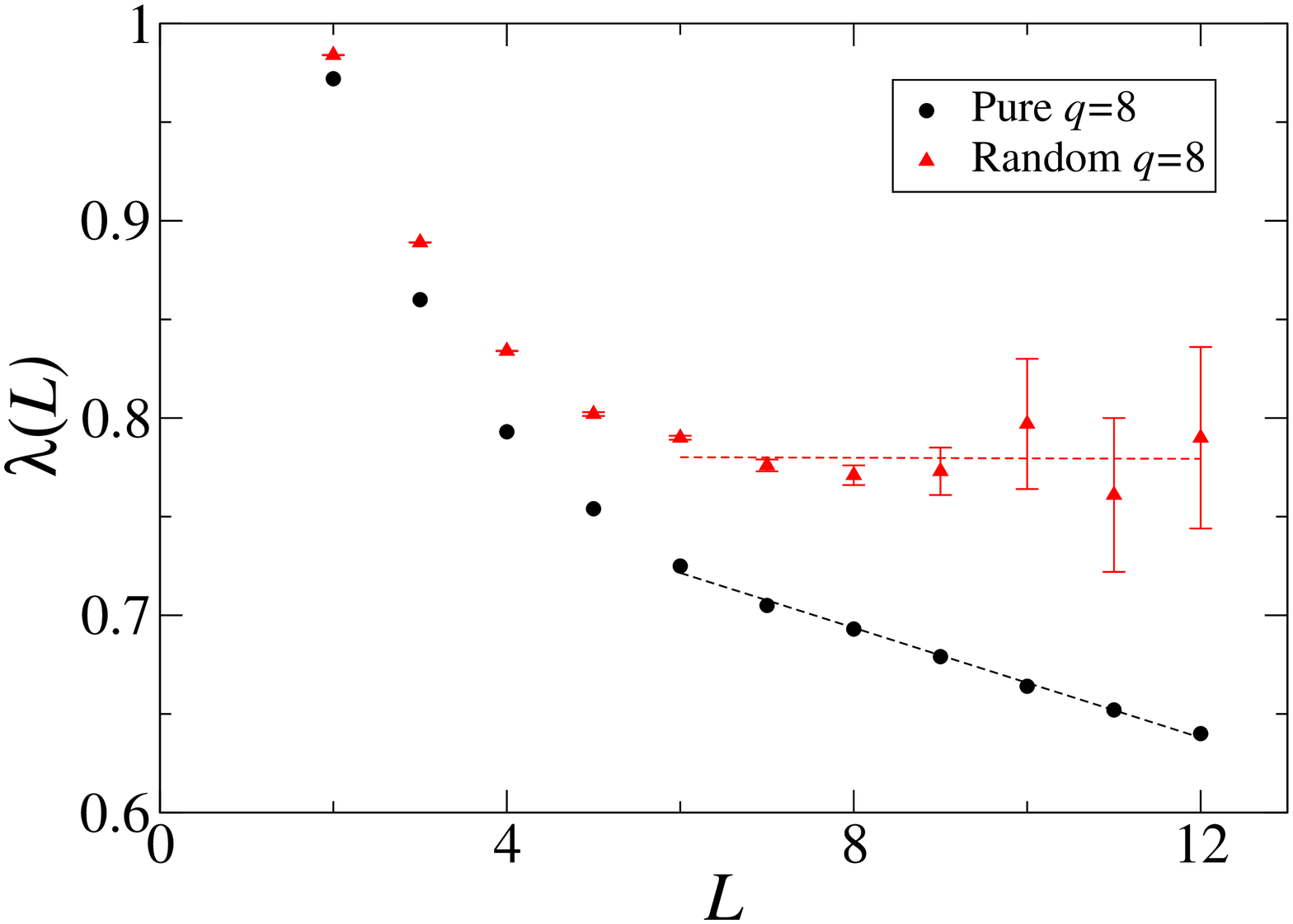}}
  \vskip 0truecm
  \caption{Evolution of the corrections to scaling to the free energy in
    a strip geometry (taken from Cardy and 
    Jacobsen~\cite{CardyJacobsen97}).} 
  \label{figCJ} 
\end{figure}

The dynamics of the Monte Carlo simulations leads to compatible 
conclusions. The energy autocorrelation time $\tau_E$ is indeed
exponentially  large (with the system size) when a non vanishing order-disorder
interface tension $\sigma_{\rm o.d.}$ exists,
$$\tau_E\sim L^{d/2}{\rm e}^{2\sigma_{\rm o.d.}L^{d-1}},$$
while it is only increasing as a power law at second-order transitions,
$$\tau_E\sim L^z,$$
with a dynamical exponent $z$ which strongly depends on the algorithm
used.
This latter situation is indeed observed~\cite{Chatelain00} 
in the disordered $8-$state
Potts model (figure~\ref{figZ}).

Cardy and Jacobsen on the other hand used transfer matrix calculations~\cite{CardyJacobsen97},
measuring the free energy density $\bar f_L$ whose corrections to scaling
behave at first-order
transitions like
$\bar f_L\sim l_\infty+O(L^{-d}{\rm e}^{-L/\xi})$. Plotting then
$\lambda(L)=\ln(\bar f_L-f_\infty)+d\ln L$ vs the strip width $L$
should give asymptotically a straight line with a slope given by the
inverse correlation length $1/\xi$. In the presence of randomness, the
curve corresponding to the $8-$state Potts model indicates a diverging 
correlation length as expected at a second-order phase transition
(figure~\ref{figCJ}).

\subsubsection{Comparison between finite-size scaling and conformal 
mappings}
\begin{figure}[h]
  \vskip1.1mm
  \epsfxsize=7cm
  \centerline{\epsfbox{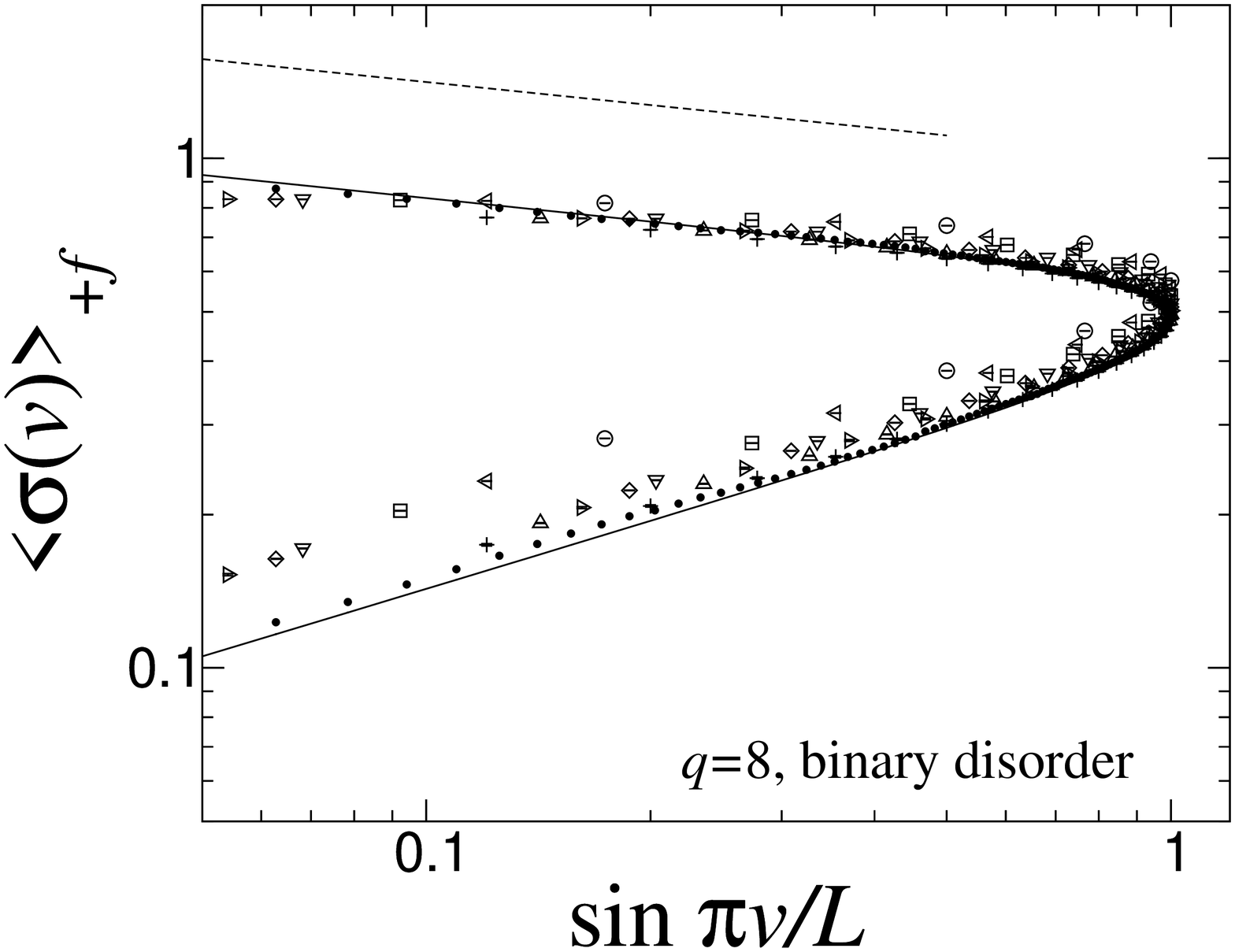}}
  \vskip 0truecm
  \caption{Fixed-free BC order parameter critical profile for the self-dual
    binary disordered $8-$state Potts model. The behaviour close to the
  fixed surface gives access to the bulk scaling dimension (the
  corresponding curve is shown in dotted line and the conformal expression
  is shown in full line while the symbols correspond to strips of various
  widths).} 
  \label{figFf} 
\end{figure}

Conformal mappings provide quite efficient techniques for the determination
of critical exponents. The validity of such an approach is nevertheless
restricted to systems where scale invariance, translation invariance
and rotation invariance do hold. This requirement is not obviously
fulfilled in random systems, since disorder breaks the symmetries.
Hopefully, one may expect that after averaging over many disorder
realizations, one is led to some effective system for which these
symmetries are restored.
This assumption can be checked from numerical simulations. Studying the
critical behaviour using an independent technique, namely finite-size-scaling
which can safely be supposed to give the right results, we then compare to
the exponents deduced from various mappings which are assumed to work.
The comparison was performed carefully in the case of the $8-$state
Potts model with binary disorder, and the technique was then applied in the
regime $q>4$, and even to asymptotically large $q$'s~\cite{JacobsenPicco00}.
The FSS results, which are considered here as the reference results, 
are shown in figure~\ref{figFSS}. In strip 
geometries, the correlation functions in the long direction and the order
parameter profile in the transverse direction (figure~\ref{figFf}) 
lead to compatible results.

\begin{figure}[ht]
  \vskip.3mm\epsfxsize=7.5cm
  \centerline{\epsfbox{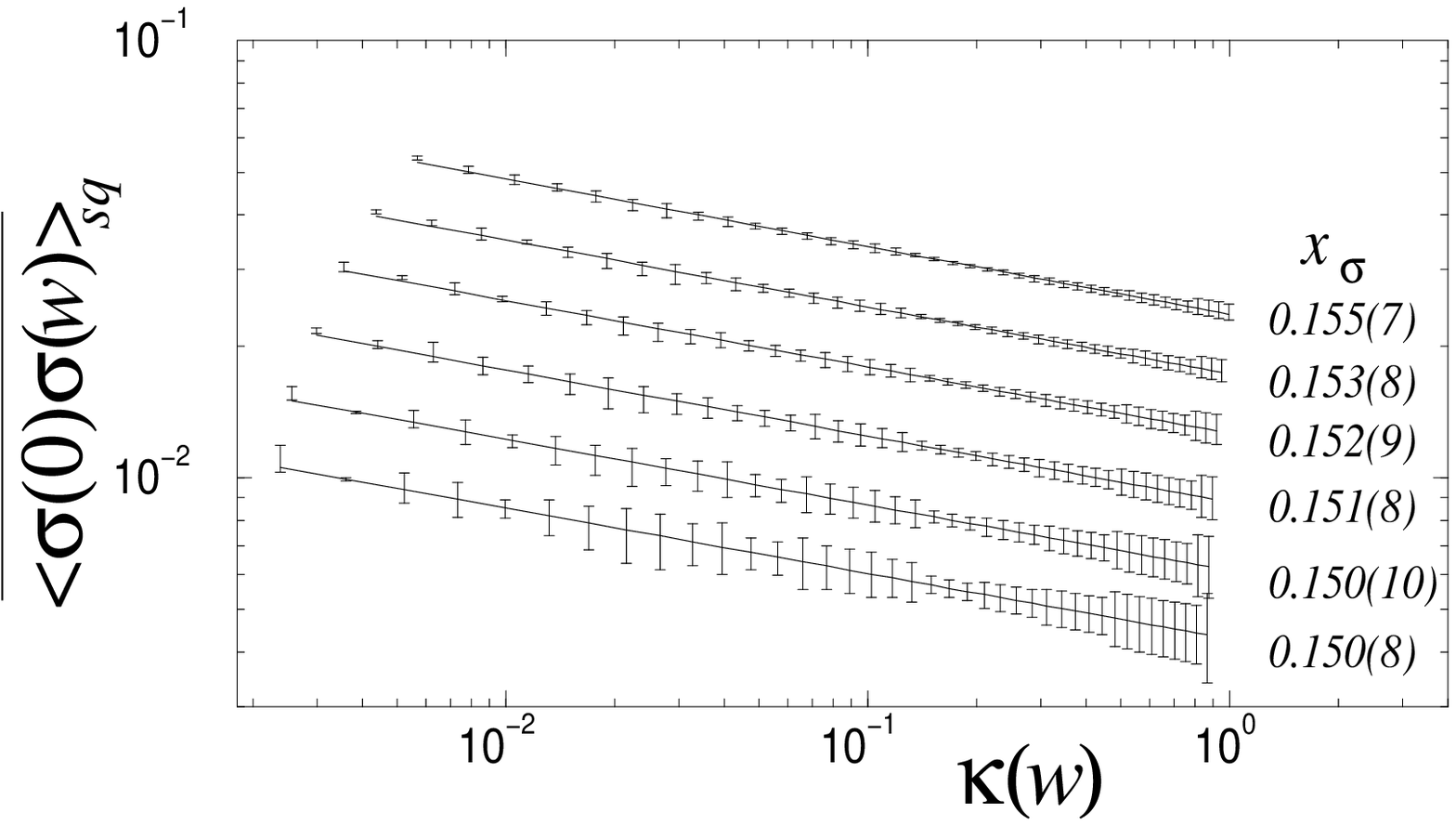}}
  \vskip.3mm\epsfxsize=8cm
  \centerline{\epsfbox{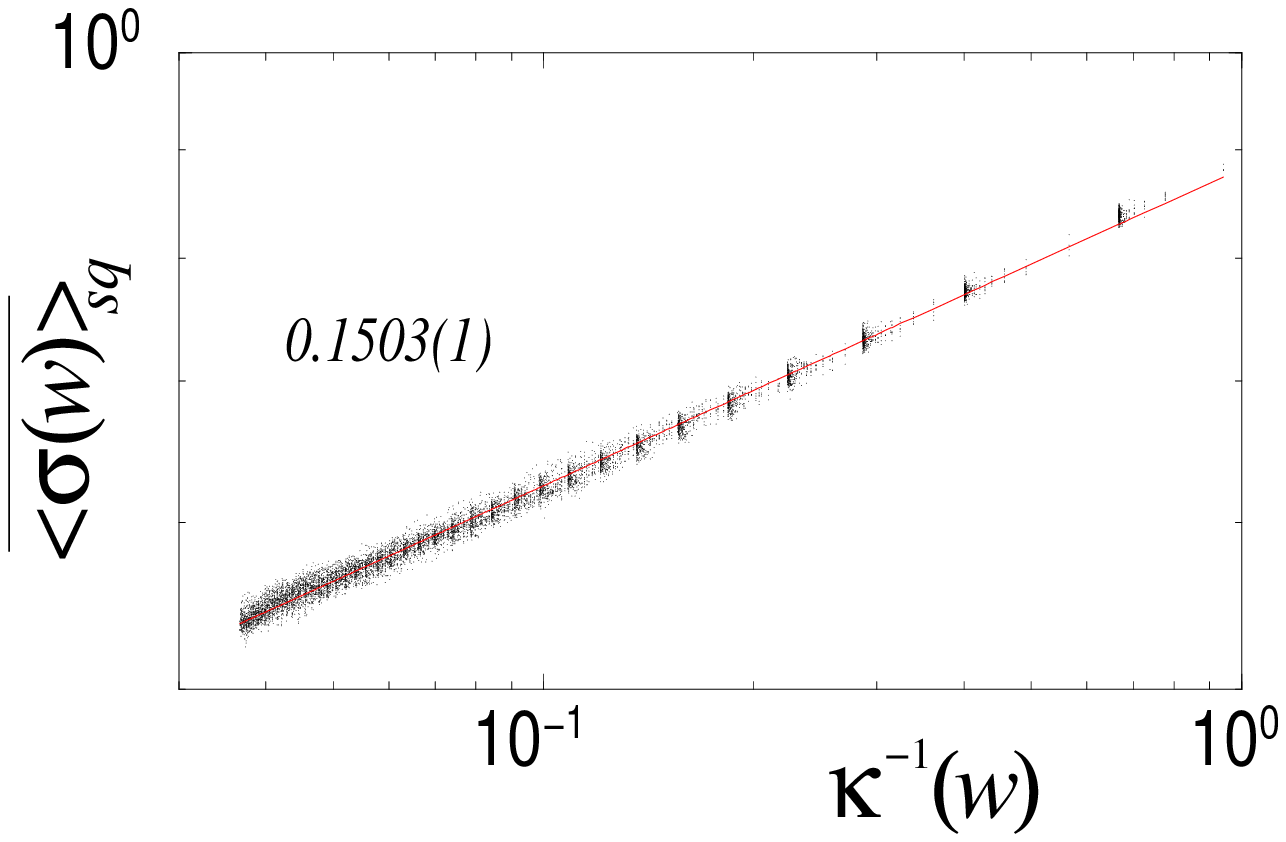}}
  \vskip 0truecm
  \caption{Fitting the correlation function  (top) 
    and the order parameter profile (bottom) in a square geometry.
    The different curves (top) correspond to different values of the
    variable $\omega$. This is the case $q=8$, with a binary distribution,
    at the optimal disorder strength estimated by the maximum of the central 
    charge in strip geometry.} 
  \label{figSqCorr} 
\end{figure}

In the square geometry, the fit of the correlation function must be
done along some curves inside the square where the scaling variable
$\omega$ remains constant, which implies that the
function $\psi(\omega)$ also remains constant and a simple
power-law fit is then needed, 
$$\overline{\langle\sigma(w_1)\sigma(w)
\rangle}_{\rm sq}\sim A_\omega |\kappa(w)|^{-x_\sigma},$$ 
where $A_\omega$
stands for the amplitude which contains $\psi(\omega)$ ($w_1$ is fixed and chosen
equal to imaginary unit $w_1={i}$). 
A difficulty occurs due
to discretization of the lattice 
Estimation of the correlation function along
the continuous curve $\omega={\rm const.}$ is done using a Taylor expansion from the data
taken at 4 neighbour points on each plaquette.
This explains the non monotonic behaviour of the error bars in 
figure~\ref{figSqCorr}. 

The case of the density profile is better to achieve,
since it involves no unknown scaling function, but a single
constant amplitude (it is a one-point correlator). All the
points inside the square enter the power-law fit
$$\overline{\langle\sigma(w)\rangle}_{\rm sq}\sim {\rm const}\times
|\kappa (w)|^{-x_\sigma}$$ 
and make it
more accurate (and there is no need of any expansion around the lattice
points). The number of points ($N^2$) being so large, we can
even forget uncertainties on each point and simply take as error bar
on the resulting exponent the standard deviation.

{\small
\begin{table}[ht]
\begin{center}
\begin{tabular}{llll}
\hline
\multicolumn{4}{c}{$x'_\sigma$ for the $8-$state Potts 
model with binary disorder} \\
\hline
Technique & Quantity & Scaling dimension & Ref. \\
\hline
\bf Standard techniques \\
$t-$dependence & $\overline{M_b(t)}$ & 0.151(1) & 
\cite{PalagyiChatelainBercheIgloi99}\\
FSS & $\overline{M_b(K_c)}$ & 0.153(1) & 
\cite{Picco98,ChatelainBerche98a}   \\
FSS & $\overline{\langle\sigma(0)\sigma(L/2)\rangle}$ & 0.159(3) & 
\cite{OlsonYoung99}   \\
Short-time dynamics & $\overline{M_b(\tau)}$ & 0.151(3) & 
\cite{YingHarada00}   \\
\hline
\bf Conformal mappings \\
Periodic strip & $\overline{\langle\sigma(0)\sigma(u)\rangle}_{\rm st}$ & 0.1505(3)
& \cite{ChatelainBerche98b,ChatelainBerche99,Chatelain00} \\
Free BC square & $\overline{\langle\sigma(0)\sigma(w)\rangle}_{\rm sq}$ & 0.152(3)
& \cite{ChatelainBerche98b,ChatelainBerche99,Chatelain00}    \\
Fixed-free strip & $\overline{\langle\sigma(v)\rangle}_{\rm st}$ & 0.150(1) 
& \cite{PalagyiChatelainBercheIgloi99,Chatelain00}   \\
Fixed BC square &  $\overline{\langle\sigma(w)\rangle}_{\rm sq}$ & 0.1503(1) 
& \cite{ChatelainBerche98b,ChatelainBerche99,Chatelain00}    \\
\hline
\end{tabular}
\end{center}
\caption{Comparison between temperature-dependence, FSS and 
short-time dynamics scaling results for the magnetic exponent 
$\beta'/\nu'$ and the scaling dimension $x'_\sigma$ at the random fixed point
deduced from the 
logarithmic and the Schwarz-Christoffel mappings 
 for the $8-$state Potts. In the first line, 
the exponent $\beta'$
deduced from the
temperature dependence is close to $x'_\sigma$, since the value of the
correlation length exponent is found very close to 1 ($\nu'\simeq 1.01(1)$).
We note also that in all these references but Ref.~\cite{OlsonYoung99},
a binary distribution of disorder was used.
\label{tab2}}
\end{table}
}

The results for the magnetic scaling dimension measured using Finite-Size
Scaling techniques, compared to the mappings onto strips or square
geometries, are compared in table~\ref{tab2} in the case of the $8-$state
Potts model with a self-dual
binary probability distribution of coupling strengths. 
First we note that the transition is second-order, even in the
regime $q>4$, as predicted by the Imry-Wortis agreement. 
More important for the following is the fact that
the agreement between different techniques is quite 
fair and leads to the conclusion that conformal techniques can be applied
here, in spite of the lack of the symmetry properties which should in principle
be required.
This is due to the fact that we are interested in {\em average quantities}. The
system thus becomes, on average, translationnaly, 
rotationally invariant, as well
as scale invariant at the critical point.
This is an important point, because the use of conformal mappings is more
accurate than standard FSS methods, and the comparison between different
schemes in the $2\le q\le 4$ regime will require a great accuracy, as already noticed in
table~\ref{tab1}.

\subsection{Regime $q\le 4$}
\subsubsection{Tests of replica symmetry}
\begin{figure}
  \vskip.3mm\epsfxsize=7.5cm
  \centerline{\epsfbox{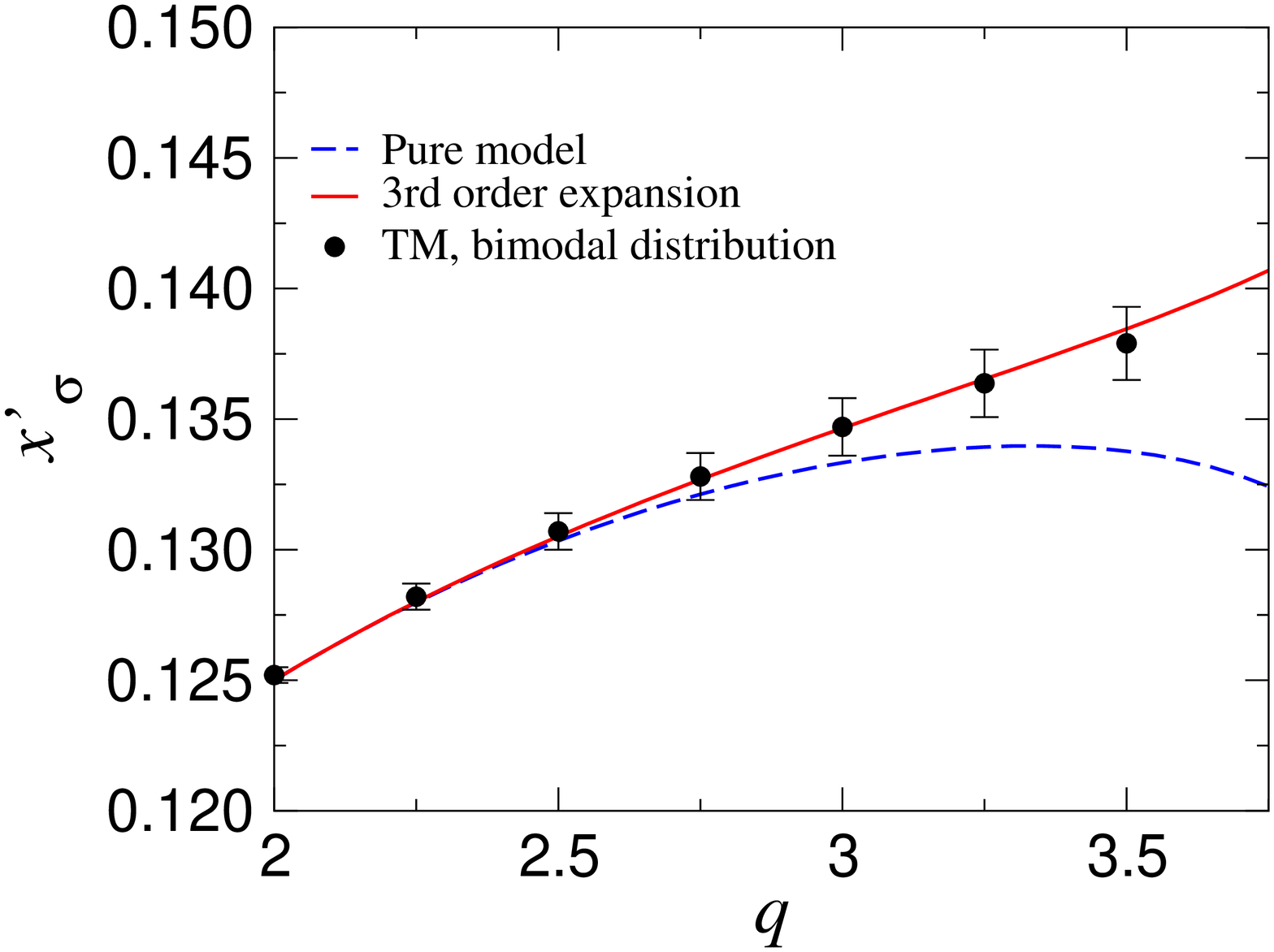}}
  \vskip 1truecm
  \vskip.3mm\epsfxsize=7.5cm
  \centerline{\epsfbox{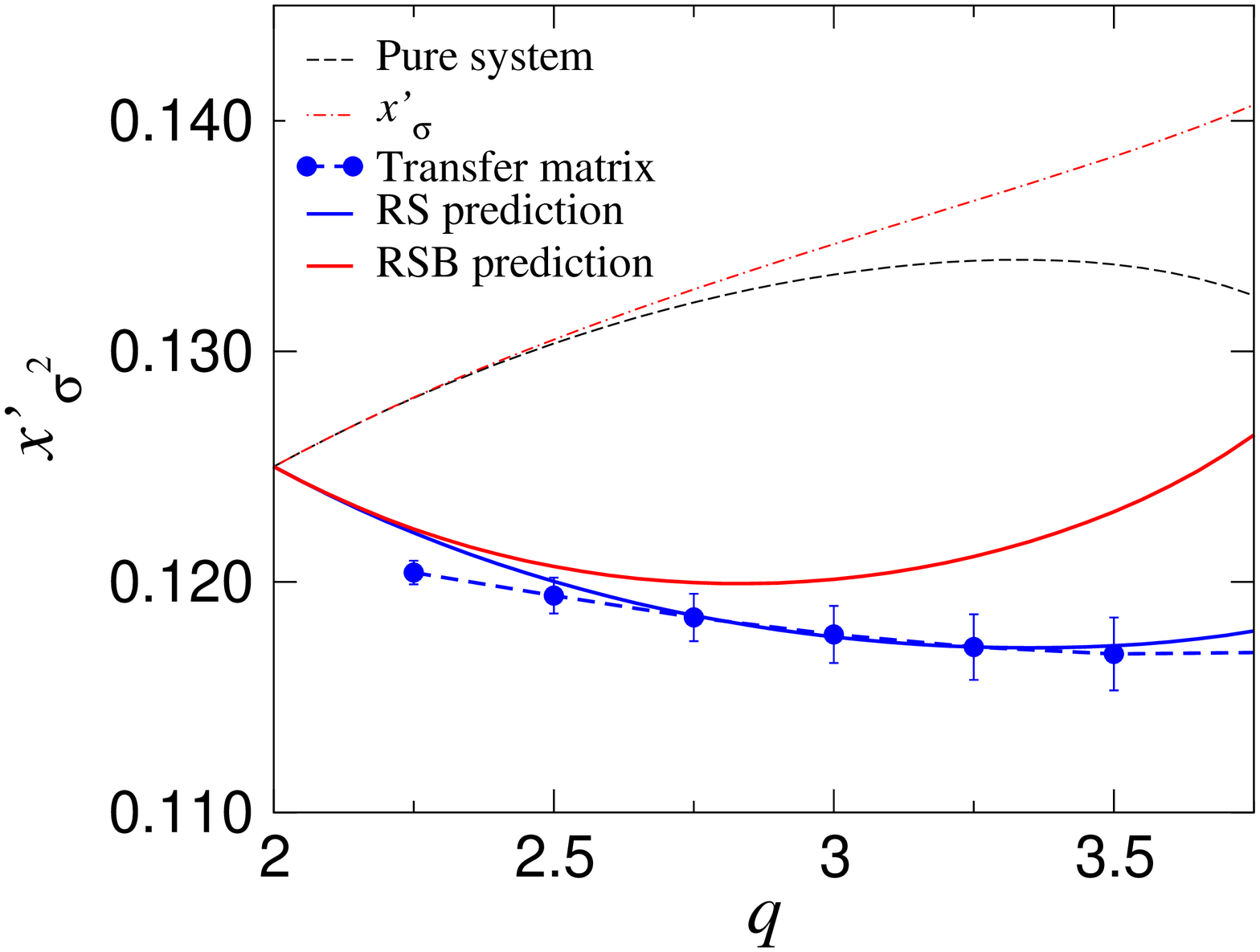}}
  \vskip 0truecm
  \caption{Top: Scaling dimension of the order parameter (binary
    disorder) compared to the third order
    expansion of Dotsenko and 
    co-workers~\protect\cite{DotsenkoPiccoPujol95a}. The scaling dimension
    corresponding to the pure model is shown for comparison.
    Bottom: Exponent of the second moment of the spin-spin correlation function
    as a function of the number of states of the disordered Potts model 
    (binary disorder). The
    comparison is done with Replica Symmetry  and Replica Symmetry Breaking
    scenarios~\protect\cite{DotsenkoDotsenkoPicco97}. The agreement
    with the RS result is quite good around $q=3$. When 
    $q$ is close to 2, the discrepancy can be attributed
    to the weak relevance of disorder. 
    We indeed used a simple exponential fit as can
    be expected at a stable disordered FP, 
    but at $q=2$, one knows from Ludwig's
    results that logarithmic corrections 
    must be added. These corrections can also
    influence the vicinity of $q=2$ in a numerical approach.} 
  \label{figxs2-vs-qbis} 
\end{figure}

First of all, the fitting procedure has to be validated.
From the exponential decay of the average spin-spin
correlation function, the exponent 
$x'_\sigma $ is deduced and presented as a function of 
$q$ for the case of a binary disorder 
in Fig.~\ref{figxs2-vs-qbis}. These results were first 
reported by Cardy and Jacobsen 
in Ref.~\cite{CardyJacobsen97}.
The agreement with the third order expansion 
in equation~(\ref{eq:x_s-dotsenko}) is extremely good
especially in the region where the expansion is supposed to be valid when $q$
is not too far from the Ising model value $q=2$. The quality of the data
confirms the reliability of the averaging 
procedure
(table~\ref{tab4}).
Even close to the marginally irrelevant case of
the Ising model where logarithmic corrections are known to be present for
some quantities, we note that the numerical data are quite satisfactorily 
in agreement with
the perturbative results. The agreement is made better
by the absence of 
logarithmic corrections for the {\it average}
correlation function at $q=2$, and we will see
that this observation is no longer true in the following study of other 
moments.

The question of a possible breaking of replica 
symmetry in disordered
systems is very controversial and far from being settled, especially
in spin glasses (see e.g 
Refs.~\cite{Mezard,MarinariParisiRuizLorenzoRitort96,MarinariEtal98a,BokilBrayDrosselMoore98,MarinariEtal98b}).
In the context of disordered Potts ferromagnets, the question was first addressed
by Dotsenko et al~\cite{DotsenkoDotsenkoPicco97}. 
In order to test between 
Replica Symmetry  and Replica Symmetry Breaking
schemes, Dotsenko et al performed a 
second order expansion of the exponent of the 
second moment of the spin-spin correlation function decay in both 
cases (Equations~(\ref{eq:RS})). 
MC simulations were first performed at $q=3$ but were not completely
conclusive, although in favour of Replica Symmetry: the perturbation expansion
leads to $x'_{\sigma^2}(3)=0.1176$ and $x''_{\sigma^2}(3)=0.1201$ according
to equations~(\ref{eq:RS}), while previous numerical results lead to
$0.113(1)$~\cite{DotsenkoDotsenkoPicco97},
$0.1140(5)$~\cite{Lewis99}, 
$0.116(1)$~\cite{PalagyiChatelainBercheIgloi99} and
$0.119(2)$~\cite{OlsonYoung99}.

{\small
\begin{table}[h]
\begin{center}
\begin{tabular}{lll}
\hline
$q$ & \multicolumn{2}{c}{$x'_\sigma$}\\
\cline{2-3}
& Expansion~(\protect\ref{eq:x_s-dotsenko}) & TM result (Ref.~\cite{ChatelainBerche00}) \\ \hline
2 & 0.12500 & 0.1252(3)\\
2.25 & 0.12800 & 0.1282(5)\\
2.5 & 0.13051 & 0.1307(7)\\
2.75 & 0.13269 &  0.1328(9)\\
3. & 0.13465 &  0.1347(11)\\
3.25 & 0.13653 &  0.1364(13)\\
3.5 & 0.13845 &  0.1379(14)\\
\hline
\end{tabular}
\end{center}
\caption{Comparison of the numerical results for the magnetic scaling dimension
  (bimodal probability distribution) $x'_\sigma$ with the third order
  expansion of Dotsenko and 
  co-workers~\protect\cite{DotsenkoPiccoPujol95a}. The 
  error bars systematically contain the analytical value.
  }
\label{tab4}
\end{table}
}
{\small
\begin{table}[h]
\begin{center}
\begin{tabular}{llll}
\hline
$q$&\multicolumn{2}{c}{Perturbative results}&TM result\\
\cline{2-4}
         & $x'_{\sigma^2}$ & $x''_{\sigma^2}$ & (Ref.~\cite{ChatelainBerche00})\\
        \cline{2-4}
2.25    &0.12213        &0.12229        &0.1204(5)      \\      
2.5     &0.12002        &0.12067        &0.1194(8)      \\      
\bf 2.75        &\bf 0.11854    &0.11997        &\bf 0.1185(10) \\      
\bf 3.  &\bf 0.11761    &0.12011        &\bf 0.1177(12) \\
\bf 3.25 &\bf 0.11718   &0.12110        &\bf 0.1172(14)\\       
\bf 3.5 &\bf 0.11723    &0.12304        &\bf 0.1169(16) \\
\hline
& $x'_\varepsilon$ & $x''_\varepsilon$ & (Ref.~\cite{Jacobsen00}) \\
\cline{2-4}
\bf 2.5 & \bf 1.006 & 1.000 & \bf 1.00(1) \\
\bf 2.75 & \bf 1.013 & 1.000 & \bf 1.01(1) \\
\bf 2.5 & \bf 1.023 & 1.000 & \bf 1.02(1) \\
\hline
\end{tabular}
\end{center}
\caption{Decay exponent of the second moment of the spin-spin correlation 
  function compared to Replica Symmetry and Replica Symmetry Breaking
  expressions of Eqs.~(\ref{eq:RS}). The results 
  written in bold face correspond 
  to the range of values of $q$ where the agreement
  is particularly satisfactory.
  The second part of the table presents Jacobsen's 
results~\cite{Jacobsen00} 
for the
  exponent of the average energy correlation function.\label{table-xs2-vs-q}}
\end{table}
}

Conclusive results for different values of $q$ were then obtained using
transfer matrices.
Close to $q=2$, the proximity of the marginally irrelevant Ising FP will
surely alter the data, as a reminiscent effect of the logarithmic 
corrections present exactly at $q=2$ for the second 
moment~\cite{Ludwig87,deQueirozStinchcombe96}. 
Too large values of
$q$ on the other hand are not very helpful in order to check perturbation
expansions which break down when one explores higher values of the expansion 
parameter. One thus has to
balance between these two extreme situations and the comparison between numerical
data and perturbation results should be conclusive around $q=3$ or slightly below. 
The TM technique thus appears to be well adapted, since it is capable to deal
with non integer values of $q$.
The comparison is shown in Fig.~\ref{figxs2-vs-qbis} for the bimodal 
probability distribution and the results are also
given in table~\ref{table-xs2-vs-q}. In the convenient domain for the test, 
around $q=3$, results are written in bold face.
 The agreement with Replica Symmetry
is quite convincing. The results for the exponent associated to the average
energy-density correlations~\cite{Jacobsen00} confirm this conclusion.


\subsubsection{Multiscaling}
The multiscaling behaviour of the spin-spin correlation functions is
noticeable in the $p-$dependent set of exponents of the reduced moments 
$$\overline{\langle\sigma(0)\sigma(R)\rangle^p}^{1/p}.$$ 
In Ref.~\cite{ChatelainBerche00}, exhaustive computation of 50 different moments in the range
$0\leq p\leq 5$ were performed
in the strip geometry, and the associated scaling dimensions followed from a
semi-log fit $\ln \overline{\langle\sigma(0)\sigma(u)\rangle^p}
$ vs $\ln u$, followed by an extrapolation
to $L\to\infty$. The numerical results were compared
to the first order expansion of Ludwig and to 
the second order
expansion in the RS scheme in Eq.~(\ref{eq:x_lewis}). The second order
result is clearly very good up to values of $p$ close to 3 and then breaks down
as already noticed by Lewis~\cite{Lewis99}. 

\begin{figure}[t]
  \vskip.5mm\epsfxsize=8cm
  \centerline{\epsfbox{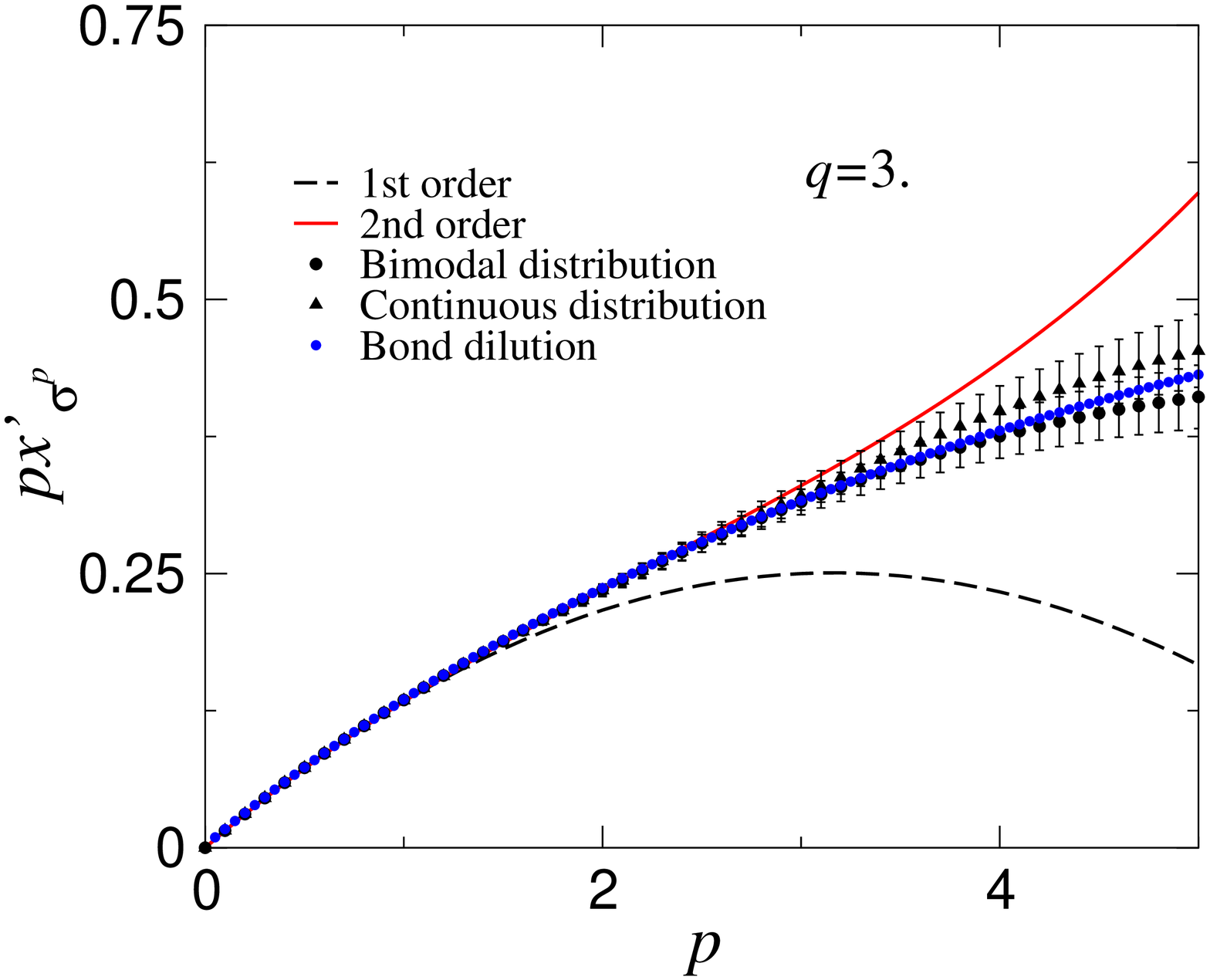}}
  \vskip 0truecm
  \caption{Comparison of the multi-fractal exponents (moment of
the correlation function $\overline{\langle\sigma(0)\sigma(u)\rangle^p}$) with the second order
expansion of Lewis in the RS scheme~\protect\cite{Lewis98} for
both the bimodal and the continuous probability distributions.} 
  \label{figq3_moments_pxbis} 
\end{figure}

An alternate presentation of the results (used e.g. by Ludwig~\cite{Ludwig90})
is given by the scaling dimension of the moment of the correlation function
itself, $\overline{\langle\sigma(0)\sigma(R)\rangle^p}$
(not the reduced function $\overline{\langle\sigma(0)\sigma(R)\rangle^p
}^{1/p}$). The scaling dimension is thus simply $px'_{\sigma^p}(q)$, hereafter
denoted by $X'_{\sigma^p}(q)$.
An example, with $q=3$, is shown in Fig.~\ref{figq3_moments_pxbis} where
we have shown the results obtained with the bimodal and the continuous 
self-dual probability distributions 
at the optimal disorder amplitude as well as the dilution case
at optimal dilution. Once again, we find a fair agreement
between the numerical data and the perturbative result which confirms 
universality, i.e. the exponent associated to a given moment of the
correlation function does not depend on the detailed probability distribution of 
the coupling strengths.

\begin{table}
\begin{center}
\begin{tabular}{lllll}
\hline
&\multicolumn{2}{c}{average}&\multicolumn{2}{c}{typical}\\
\cline{2-5}
$q$ & $x'_{\varepsilon^1}$ & TM & $x'_{\varepsilon^0}$ & TM \\
\hline
2.5 & 1.006 & 1.00(1) & 1.023 & 1.02(1) \\
2.75 & 1.013 & 1.01(1) & 1.051 & 1.04(2) \\
3. & 1.023 & 1.02(1) & 1.090 & 1.06(3) \\
\hline
\end{tabular}
\end{center}
\caption{Decay exponent of the average and typical energy-energy
  correlation functions: comparison between perturbative expansion from Ref.~\cite{JengLudwig01} and
transfer matrix computation
(from Jacobsen~\cite{Jacobsen00}).\label{tab6}}
\end{table}

\subsubsection{Probability distribution of correlation functions}
\begin{figure}[ht]
  \vskip.3mm\epsfxsize=10cm
  \centerline{\epsfbox{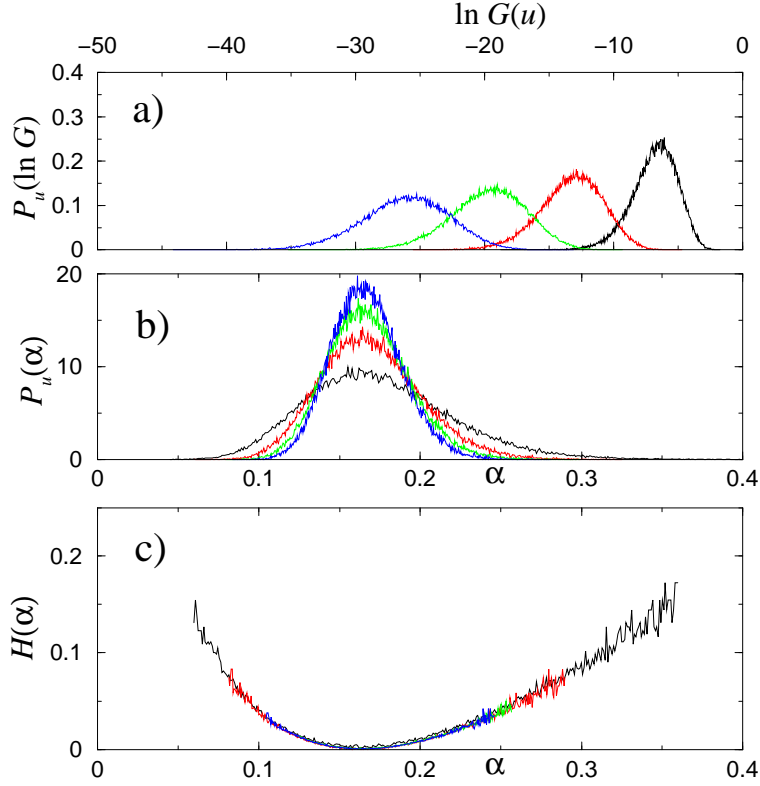}}
  \vskip 0truecm
  \caption{Fitting the probability distribution of the spin-spin
    correlation function to get a collapse onto a universal multiscaling
    function.} 
  \label{DistProbG_q3Fig} 
\end{figure}

In Ref.~\cite{Ludwig90}, Ludwig presented a remarquable discussion of
the spin-spin correlation function probability distribution.
 The relevant
information on the large distance behaviour is 
encoded in the multiscaling function $H(\alpha)$, which is simply
the Legendre transform of the set of independent scaling indexes 
$X'_{\sigma^p}(q)$. Setting ${\rm d}X'_{\sigma^p}(q)=\alpha{\rm d}p$, this
function is simply obtained by 
$$H(\alpha)=X'_{\sigma^p}(q)-\alpha p.$$ 
The
geometrical
interpretation of this Legendre transform follows from the relation
$\frac{\partial H}{\partial \alpha}=-p$ where $\alpha$ is defined by
$\frac{\partial X'_{\sigma^p}(q)}{\partial p}=\alpha$. The scaling
dimension $x'_{\sigma^p}(q)$ is obtained on the plot of $H(\alpha)$ 
by the intercept of the tangent of slope
$-p$ with the abscissa axis. An example of multiscaling function $H(\alpha)$
deduced from the numerical data with the bimodal probability distribution 
is shown in Fig.~\ref{DistProbG_q3Fig}.

In this section, we follow Ludwig's arguments and report a
numerical study of the correlation function probability distribution in the 
cylinder geometry.

According to the results of the previous section, the moments of the
spin-spin correlation 
function along the strip asymptotically behaves as follows:
\begin{equation}
        \overline{G^p(u)}\equiv\overline{\langle\sigma(0)\sigma(u)\rangle^p}
        \sim B_p{\rm e}^{-\frac{2\pi u}{L}X'_{\sigma^p}}
\label{eqProb1}
\end{equation}
and are defined in terms of the probability distribution ${\cal P}[G(u)]$:
\begin{equation}
        \overline{G^p(u)}=\int_0^1{\rm d} G(u){\cal P}[G(u)]G^p(u).
\label{eqProb2}
\end{equation}
Following Ludwig, we introduce the variable $Y(u)=-\ln G(u)$
and write $G^p(u)={\rm e}^{-pY(u)}$. Using the identity
${\cal P}[G(u)]{\rm d}G={\cal P}[Y(u)]{\rm d}Y$ and 
equations~(\ref{eqProb1}) and (\ref{eqProb2}), one obtains
$$        
        \int_0^\infty{\rm d} Y(u){\cal P}[Y(u)]{\rm e}^{-pY(u)}
        \sim B_p{\rm e}^{-\frac{2\pi u}{L}X'_{\sigma^p}}
        \label{eqProb3}
$$
which leads to
the expression of the probability distribution by inverting the 
Laplace transform ($\delta>0$):
$$
        {\cal P}[Y(u)]=\frac{1}{2{i}\pi}\lim_{\delta\to 0}
        \int_{\delta-{i}\infty}^{\delta+{i}\infty} 
        {\rm d} pB_p{\rm e}^{-\frac{2\pi u}{L}\left[X'_{\sigma^p}
        -\frac{Y(u)}{2\pi u/L}p\right]}.
$$
The amplitude $B_p$ is assumed to be smoothly dependent on $p$
(this can be checked numerically), and its dependence  
 can be forgotten with respect to the exponential, since it only introduces a correction 
 when $2\pi u/L\to\infty$.
Let us define the function $h(p)=X'_{\sigma^p}
        -\frac{Y(u)}{2\pi u/L}p$.
In the large distance limit $2\pi u/L\to\infty$, the integral can be 
evaluated by the saddle-point approximation at the minimum $p_0$ of $h(p)$:
$$
        \left(\frac{\partial}{\partial p}X'_{\sigma^p}\right)_{p_0}=
        \frac{Y(u)}{2\pi u/L}
$$
Instead of $Y(u)$, we define the scaled variable 
$\alpha=\frac{Y(u)}{2\pi u/L}$, and the saddle point value at $p_0$ only depends
on this variable $h(p_0)=H(\alpha)$. 
We thus obtain the probability distribution
\begin{equation}
        {\cal P}[Y(u)]\sim\exp\left[{-\frac{2\pi u}{L}H\left(\frac{Y(u)}
        {2\pi u/L}\right)}\right],
\label{eqProb6}
\end{equation}
or, using ${\cal P}[Y(u)]{\rm d}Y={\cal P}(\alpha){\rm d}\alpha$, 
\begin{equation}
        {\cal P}(\alpha)\sim\frac{2\pi u}{L}\exp\left[{-\frac{2\pi u}
        {L}H(\alpha)}\right].
\label{eqProb7}
\end{equation}

The multi-fractal function contains the essential information on the probability
distribution. 
A correction to the leading behaviour given by the saddle-point 
approximation is needed here to improve the data collapse onto a
single multiscaling function.
If we expand the function $h(p)$ close to $p_0$, $h(p)\simeq H(\alpha)+\frac{1}
{2}h''(p_0)(p-p_0)^2$, with $h''(p_0)>0$ we obtain, instead of Eq.~(\ref{eqProb6}),
the following result for 
the probability 
distribution ${\cal P}[Y(u)]$~\cite{Debruijn}:
\begin{equation}
        {\cal P}[Y(u)]\sim\left(\frac{2\pi u}{L}\right)^{-1/2}
        \exp\left[{-\frac{2\pi u}
        {L}H
        \left(\frac{Y(u)}
        {2\pi u/L}\right)}\right]
\label{eqProb9}
\end{equation}
and a correction appears in ${\cal P}(\alpha)$:
\begin{equation}
        {\cal P}(\alpha)\sim\left(\frac{2\pi u}{L}\right)^{1/2}
\exp\left[{-\frac{2\pi u}
        {L}H(\alpha)}\right].
\label{eqProb10}
\end{equation}
which enables to extract
$H(\alpha)$ at fixed $\alpha$  by fitting the probability 
distribution to
the expression
\begin{equation}
  \ln{\cal P}(\alpha)-\frac{1}{2}\ln\frac{2\pi u}{L}={\rm const}-\frac{2\pi u}
        {L}H
        (\alpha).
\label{eqProb8}
\end{equation}
It is shown in Fig.~\ref{DistProbG_q3Fig} where the probability
distribution of the spin-spin correlation function was obtained
after collecting the results over $96\ 000$ disorder realizations
in 50 classes~\cite{ChatelainBerche00}. All the data collapse onto a single
multiscaling function $H(\alpha)$, for different distances $u$ and this
is even the case with still good accuracy for
different strip widths or different probability 
distributions~\cite{ChatelainBerche00,ChatelainBerche01}.


\section{Conclusion and summary of the main results}
The two-dimensional Potts model is the ideal framework to test the influence
of quenched randomness on phase transitions and critical behaviour.
It exhibits a second-order phase transition completely characterised
by conformal invariance when the number of states per spin is lower or 
equal to 4 and a first-order transition above. The transition line is
exactly known, and it is easy to build in the random case
probability distributions of coupling strengths which preserve the
self-duality relation.

With respect to these advantages, many  results concerning the
effect of a weak disorder were obtained
during the last decade using perturbations expansions around the pure fixed
point. Different solutions were considered, first replica symmetric
solutions where the symmetry between all the replicas is supposed to be
preserved in the renormalization equations, then the case of a 
spontaneous breaking of the replica symmetry was also studied
perturbatively.

Numerical studies were also performed from different sides. 
First of all, Monte Carlo simulations coupled to finite-size scaling analysis,
then transfer matrices and sophisticated graph and loop algorithms 
coupled to extensive use of conformal mappings
in order to extract the values of the critical exponents with a 
pretty good accuracy.
All the results were in a fair agreement with the perturbative
expansions close to the Ising model limit and concluded
in favour of replica symmetric scenarios. The multiscaling 
properties of the order-parameter and energy density were also analysed,
and a characterisation of the probability distributions of the 
correlation functions was made in terms of universal
multiscaling functions.
The regime $q>4$ where the pure model exhibits a first order transition was also 
extensively studied, but did not display any particular features compared to 
the regime $q\le 4$ in the presence quenched randomness. 

The most important question which remains open up to now is the identification
of the conformal field theories which could describe the random fixed point.
This is a delicate program, since in the replica limit of coupled models,
the theory is not unitary (with 
a central charge which evolves continuously with the value of $q$).
We may hope some progress in this direction for the near future, which would 
undoubtedly achieve a considerable progress in the understanding of two-dimensional
disordered systems.

\vspace{10mm}\noindent{\bf Acknowledgement}

The material discussed here was partly presented at the Helsinki
SPHINX ESF Conference 2002 on {\it Disordered systems at low 
temperatures and their topological properties}, and
at the {\it Ising lectures} 2002 at Lviv. BB greatly acknowledges
the organisers of the Helsinki workshop, M. Alava and H. Rieger and of
the Lviv lectures, Y. Holovatch, I. Mryglod, and K. Tabunshchyk, for
the opportunity given to present this review.

\end{document}